\documentclass[aps,prd,twocolumn,superscriptaddress,showpacs]{revtex4}
\usepackage{epsfig}
\usepackage{graphicx}
\usepackage{dcolumn}
\usepackage{amsmath}
\usepackage{subfigure}
\usepackage{xspace}
\usepackage[usenames,dvips]{pstcol}
\usepackage{color}
\usepackage{amssymb}
\usepackage{epsfig}
\usepackage{footnote}
\usepackage{longtable}
\usepackage{fancyhdr}
\usepackage{subfigure}
\usepackage{xspace}

\newcommand{\ppbar}{p{\bar p}}
\newcommand{\invpb}{\rm pb^{-1}} % Should be roman in PRL
\newcommand{\bfinvpb}{\bf pb^{-1}} % Should be roman in PRL
\newcommand{\roots}{{\sqrt s}}
 
\newcommand{\Et}{E_T}
\newcommand{\Pt}{p_T}
\newcommand{\Ht}{H_T}
\newcommand{\egg}{e\gamma\gamma}
\newcommand{\mugg}{\mu\gamma\gamma}
\newcommand{\eeggmet}{ee\gamma\gamma\met}
\newcommand{\llggmet}{\ell\ell\gamma\gamma\met}
\newcommand{\lgal}{\ell\gamma}
\newcommand{\lgX}{\ell\gamma\plus X}
\newcommand{\egX}{e\gamma\plus X}
\newcommand{\mugX}{\mu\gamma\plus X}
\newcommand{\ggX}{\gamma\gamma+X}
\newcommand{\emugX}{e\mu\gamma+X}

\def\lum{{\cal L}}
\newcommand{\pmasym}[2]{^{+#1}_{-#2}}
\newcommand{\gt}{>}

\newcommand{\met}{{\not\!\!E}_{T}}
\newcommand{\metvec}{{\not\!\! \vec{E}_T}}
\newcommand{\lepvec}{{\vec{E}_T^\ell}}
\newcommand{\phovec}{{\vec{E}_T^\gamma}}

\newcommand{\lgmet}{\ell\gamma\met}
\newcommand{\egmet}{e\gamma\met}
\newcommand{\eg}{e\gamma}
\newcommand{\mug}{\mu\gamma}
\newcommand{\mugmet}{\mu\gamma\met}
\newcommand{\llg}{\ell\ell\gamma}
\def\ttg{t\bar{t}\gamma}
\newcommand{\lgg}{\ell\gamma\gamma}

\newcommand{\eeg}{ee\gamma}
\newcommand{\mumug}{\mu\mu\gamma}
\newcommand{\Wenu}{W \rightarrow e\nu}
\newcommand{\Wmunu}{W \rightarrow \mu\nu}
\newcommand{\Wg}{W \gamma}
\newcommand{\Wgg}{W \gamma\gamma}
\newcommand{\Zee}{Z \rightarrow e^+e^-}

\newcommand{\Zg}{Z \gamma}
\newcommand{\Zgg}{Z \gamma\gamma}
\newcommand{\Z}{Z}
\newcommand{\Zgstar}{Z\kern -0.25em/\kern -0.15em\gamma^*}
\newcommand{\Zmumu}{Z \rightarrow \mu^+\mu^-}

\newcommand{\runonelumi}{86}
\newcommand{\goes}{\kern -0.18em\rightarrow\kern -0.18em}
\newcommand{\plus}{\kern -0.18em +\kern -0.18em}
\newcommand{\degs}{\mbox{$^{\circ}$}}
\newcommand{\MeV}{\ensuremath{\mathrm{\ Me\kern -0.1em V}}\xspace}
\newcommand{\MeVc}{\ensuremath{\mathrm{\ Me\kern -0.1em V\kern -0.1em 
/\mathit{c}}}\xspace}
\newcommand{\MeVcsq}{\ensuremath{\mathrm{\ Me\kern -0.1em V\kern -0.1em 
/\mathit{c}^2}}\xspace}
\newcommand{\GeV}{\ensuremath{\mathrm{Ge\kern -0.1em V}}\xspace}
\newcommand{\GeVc}{\ensuremath{\mathrm{\ Ge\kern -0.1em V\kern -0.1em 
/\mathit{c}}}\xspace}
\newcommand{\GeVcsq}{\ensuremath{\mathrm{\ Ge\kern -0.1em V\kern -0.1em 
/\mathit{c}^2}}\xspace}
\newcommand{\TeV}{\ensuremath{\mathrm{Te\kern -0.1em V}}\xspace}
\newcommand{\bfTeV}{\ensuremath{\bf{Te\kern -0.1em V}}\xspace}
\newcommand{\nsGeV}{\ensuremath{\mathrm{Ge\kern -0.1em V}}\xspace}
\newcommand{\nsGeVc}{\ensuremath{\mathrm{Ge\kern -0.1em V\kern -0.1em 
/\mathit{c}}}\xspace}
\newcommand{\Etgamma}{\ensuremath{\mathrm{E_T^{\gamma}}}}
\newcommand{\Etlepton}{\ensuremath{\mathrm{E_T^{\ell}}}}

\input{total.summary}

\begin{document}
%\pagewiselinenumbers
\title{Search for New Physics in Lepton + Photon + X
 Events with $\bf\lumi$ $\bfinvpb$ of $\bf\ppbar$ Collisions at $\bf\roots$= 1.96 $\bfTeV$}
\affiliation{Institute of Physics, Academia Sinica, Taipei, Taiwan 11529, Republic of China} 
\affiliation{Argonne National Laboratory, Argonne, Illinois 60439} 
\affiliation{Institut de Fisica d'Altes Energies, Universitat Autonoma de Barcelona, E-08193, Bellaterra (Barcelona), Spain} 
\affiliation{Baylor University, Waco, Texas  76798} 
\affiliation{Istituto Nazionale di Fisica Nucleare, University of Bologna, I-40127 Bologna, Italy} 
\affiliation{Brandeis University, Waltham, Massachusetts 02254} 
\affiliation{University of California, Davis, Davis, California  95616} 
\affiliation{University of California, Los Angeles, Los Angeles, California  90024} 
\affiliation{University of California, San Diego, La Jolla, California  92093} 
\affiliation{University of California, Santa Barbara, Santa Barbara, California 93106} 
\affiliation{Instituto de Fisica de Cantabria, CSIC-University of Cantabria, 39005 Santander, Spain} 
\affiliation{Carnegie Mellon University, Pittsburgh, PA  15213} 
\affiliation{Enrico Fermi Institute, University of Chicago, Chicago, Illinois 60637} 
\affiliation{Comenius University, 842 48 Bratislava, Slovakia; Institute of Experimental Physics, 040 01 Kosice, Slovakia} 
\affiliation{Joint Institute for Nuclear Research, RU-141980 Dubna, Russia} 
\affiliation{Duke University, Durham, North Carolina  27708} 
\affiliation{Fermi National Accelerator Laboratory, Batavia, Illinois 60510} 
\affiliation{University of Florida, Gainesville, Florida  32611} 
\affiliation{Laboratori Nazionali di Frascati, Istituto Nazionale di Fisica Nucleare, I-00044 Frascati, Italy} 
\affiliation{University of Geneva, CH-1211 Geneva 4, Switzerland} 
\affiliation{Glasgow University, Glasgow G12 8QQ, United Kingdom} 
\affiliation{Harvard University, Cambridge, Massachusetts 02138} 
\affiliation{Division of High Energy Physics, Department of Physics, University of Helsinki and Helsinki Institute of Physics, FIN-00014, Helsinki, Finland} 
\affiliation{University of Illinois, Urbana, Illinois 61801} 
\affiliation{The Johns Hopkins University, Baltimore, Maryland 21218} 
\affiliation{Institut f\"{u}r Experimentelle Kernphysik, Universit\"{a}t Karlsruhe, 76128 Karlsruhe, Germany} 
\affiliation{High Energy Accelerator Research Organization (KEK), Tsukuba, Ibaraki 305, Japan} 
\affiliation{Center for High Energy Physics: Kyungpook National University, Taegu 702-701, Korea; Seoul National University, Seoul 151-742, Korea; and SungKyunKwan University, Suwon 440-746, Korea} 
\affiliation{Ernest Orlando Lawrence Berkeley National Laboratory, Berkeley, California 94720} 
\affiliation{University of Liverpool, Liverpool L69 7ZE, United Kingdom} 
\affiliation{University College London, London WC1E 6BT, United Kingdom} 
\affiliation{Centro de Investigaciones Energeticas Medioambientales y Tecnologicas, E-28040 Madrid, Spain} 
\affiliation{Massachusetts Institute of Technology, Cambridge, Massachusetts  02139} 
\affiliation{Institute of Particle Physics: McGill University, Montr\'{e}al, Canada H3A~2T8; and University of Toronto, Toronto, Canada M5S~1A7} 
\affiliation{University of Michigan, Ann Arbor, Michigan 48109} 
\affiliation{Michigan State University, East Lansing, Michigan  48824} 
\affiliation{Institution for Theoretical and Experimental Physics, ITEP, Moscow 117259, Russia} 
\affiliation{University of New Mexico, Albuquerque, New Mexico 87131} 
\affiliation{Northwestern University, Evanston, Illinois  60208} 
\affiliation{The Ohio State University, Columbus, Ohio  43210} 
\affiliation{Okayama University, Okayama 700-8530, Japan} 
\affiliation{Osaka City University, Osaka 588, Japan} 
\affiliation{University of Oxford, Oxford OX1 3RH, United Kingdom} 
\affiliation{University of Padova, Istituto Nazionale di Fisica Nucleare, Sezione di Padova-Trento, I-35131 Padova, Italy} 
\affiliation{LPNHE, Universite Pierre et Marie Curie/IN2P3-CNRS, UMR7585, Paris, F-75252 France} 
\affiliation{University of Pennsylvania, Philadelphia, Pennsylvania 19104} 
\affiliation{Istituto Nazionale di Fisica Nucleare Pisa, Universities of Pisa, Siena and Scuola Normale Superiore, I-56127 Pisa, Italy} 
\affiliation{University of Pittsburgh, Pittsburgh, Pennsylvania 15260} 
\affiliation{Purdue University, West Lafayette, Indiana 47907} 
\affiliation{University of Rochester, Rochester, New York 14627} 
\affiliation{The Rockefeller University, New York, New York 10021} 
\affiliation{Istituto Nazionale di Fisica Nucleare, Sezione di Roma 1, University of Rome ``La Sapienza," I-00185 Roma, Italy} 
\affiliation{Rutgers University, Piscataway, New Jersey 08855} 
\affiliation{Texas A\&M University, College Station, Texas 77843} 
\affiliation{Istituto Nazionale di Fisica Nucleare, University of Trieste/\ Udine, Italy} 
\affiliation{University of Tsukuba, Tsukuba, Ibaraki 305, Japan} 
\affiliation{Tufts University, Medford, Massachusetts 02155} 
\affiliation{Waseda University, Tokyo 169, Japan} 
\affiliation{Wayne State University, Detroit, Michigan  48201} 
\affiliation{University of Wisconsin, Madison, Wisconsin 53706} 
\affiliation{Yale University, New Haven, Connecticut 06520} 
\author{A.~Abulencia}
\affiliation{University of Illinois, Urbana, Illinois 61801}
\author{J.~Adelman}
\affiliation{Enrico Fermi Institute, University of Chicago, Chicago, Illinois 60637}
\author{T.~Affolder}
\affiliation{University of California, Santa Barbara, Santa Barbara, California 93106}
\author{T.~Akimoto}
\affiliation{University of Tsukuba, Tsukuba, Ibaraki 305, Japan}
\author{M.G.~Albrow}
\affiliation{Fermi National Accelerator Laboratory, Batavia, Illinois 60510}
\author{D.~Ambrose}
\affiliation{Fermi National Accelerator Laboratory, Batavia, Illinois 60510}
\author{S.~Amerio}
\affiliation{University of Padova, Istituto Nazionale di Fisica Nucleare, Sezione di Padova-Trento, I-35131 Padova, Italy}
\author{D.~Amidei}
\affiliation{University of Michigan, Ann Arbor, Michigan 48109}
\author{A.~Anastassov}
\affiliation{Rutgers University, Piscataway, New Jersey 08855}
\author{K.~Anikeev}
\affiliation{Fermi National Accelerator Laboratory, Batavia, Illinois 60510}
% ...
\collaboration{CDF Collaboration\footnote{With visitors from $^a$University of Athens, 
$^b$University of Bristol, 
$^c$University Libre de Bruxelles, 
$^d$Cornell University, 
$^e$University of Cyprus, 
$^f$University of Dublin, 
$^g$University of Edinburgh, 
$^h$University of Heidelberg, 
$^i$Universidad Iberoamericana, 
$^j$University of Manchester, 
$^k$Nagasaki Institute of Applied Science, 
$^l$University de Oviedo, 
$^m$University of London, Queen Mary and Westfield College, 
$^n$Texas Tech University, 
$^o$IFIC(CSIC-Universitat de Valencia), 
}}
\noaffiliation

%
%\collaboration{CDF Collaboration}
%\preprint{FERMILAB-PUB-06/yyy-E}
%\preprint{EFI-06/yyy}
%\preprint{EFI-05-zzz}

%\input{cdf_auth.tex}
%\input{run1_revtex4_auth.tex}

\date{\today}

%(CDF Collaboration)

\begin{abstract}
We present results of a search at CDF in $\lumi\pm\dlumi$ $\invpb$ of
$\ppbar$ collisions at 1.96 $\TeV$ for the anomalous production of
events containing a high-transverse momentum charged lepton ($\ell$,
either $e$ or $\mu$) and photon ($\gamma$), accompanied by missing
transverse energy ($\met$), and/or additional leptons and photons, and
jets (X). We use the same selection criteria as in a previous CDF Run
I search, but with an order-magnitude larger data set, a higher
$\ppbar$ collision energy, and the CDF II detector. We find
$\noflglgmet$ $\lgmet+X$ events, compared to an expectation of
$\smnoflglgmet \pm \totdsysnoflglgmet$ events. We observe
$\noflgmultil$ $\llg + X$ events, compared to an expectation of
$\smnoflgmultil \pm \totdsysnoflgmultil$ events. We find no events
similar to the Run I $\eeggmet$ event.
\end{abstract}

% PACS, the Physics and Astronomy Classification Scheme.

\maketitle

\section{Introduction}
\label{introduction}

An important test of the standard model (SM) of particle
physics~\cite{SM} is to measure and understand the properties of the
highest momentum-transfer particle collisions, which correspond to
measurements at the shortest distances.  The chief predictions of the
SM for these collisions are the numbers and types of the fundamental
fermions and gauge bosons that are produced, and their associated
kinematic distributions.  The predicted high energy behavior of the
SM, however, becomes unphysical at an interaction energy on the order
of several TeV.  New physical phenomena are required to ameliorate
this high-energy behavior.  These unknown phenomena may involve new
elementary particles, new fundamental forces, and/or a modification of
space-time geometry.  These new phenomena are likely to show up as an
anomalous production rate of a combination of the known fundamental
particles, including those detector-based signatures such as missing
transverse energy ($\met$) or penetrating particle tracks that within
the confines of the SM are associated with neutrinos and muons,
respectively.

The unknown nature of possible new phenomena in the energy range
accessible at the Tevatron is the motivation for a search strategy
that does not focus on a single model or class of models of new
physics, but presents a wide net for new phenomena. In this paper we
present the results of a comparison of standard model predictions with
the rates measured at the Tevatron with the CDF detector for final
states containing at least one high-$\Pt$ lepton (e or $\mu$) and photon,
plus other detected objects (leptons, photons, jets, $\met$).

The initial motivation for such an inclusive search (``signature-based
search'') came from the observation in 1995 by the CDF
experiment~\cite{Toback_all} of an event consistent with the
production of two energetic photons, two energetic electrons, and
large missing transverse energy $\met$~\cite{EtPt}. This signature is
predicted to be very rare in the SM, with the dominant contribution
being from the production of four gauge bosons: two W bosons and two
photons. The event raised theoretical interest, however, as it had, in
addition to large missing transverse momentum, very high total
transverse energy, and a pattern of widely-separated leptons and
photons that was consistent with the decay of a pair of new heavy
particles. 

There are many models of new physics that could produce such a
signature~\cite{lhc_wkgp}.  Gauge-mediated models of
supersymmetry~\cite{susy_gauge}, in which the lightest super-partner
(LSP) is a light gravitino, provide a model in which each partner of a
pair of supersymmetric particles produced in a $\ppbar$ interaction
decays in a chain that leads to a produced gravitino, visible as
$\met$. If the next-to-lightest neutralino (NLSP) has a photino
component, each chain also can result in a photon. Models of
supersymmetry in which the symmetry breaking is due to gravity also
can produce decay chains with photons~\cite{susy_gravity}. For
example, if the NLSP is largely photino-like, and the lightest is
largely higgsino, decays of the former to the latter will involve the
emission of a photon~\cite{kane_loop}. More generally, pair-production
of selectrons or gauginos can result in final-states with large
$\met$, two photons and two leptons.  Models with additional space
dimensions~\cite{LED} predict excited states of the known standard
model particles.  The production of a pair of excited
electrons~\cite{excited_electron} would provide a natural source for
two photons and two electrons (although not $\met$ unless the pair
were produced with some other, undetected, particle.). As in the case
of supersymmetry, there are many parameters in such models, with a
resulting broad range of possible signatures with multiple gauge
bosons~\cite{Geraldine}.

Rather than search the huge parameter space of the models current at
that time, the CDF Run I analyses that followed up on the $\eeggmet$
event used a strategy of ``signature-based'' inclusive searches to
cast a wider net for new phenomena: in this case one search for two
photons + X ($\ggX$)~\cite{Toback_all}, and a second for one lepton +
one photon + X ($\lgX$)~\cite{Jeff_PRD,Jeff_PRL,Jeff_thesis}, where X
can be $e$, $\mu$, $\gamma$, or $\met$, plus any number of jets. In
particular the latter signature, the subject of this present paper,
would be sensitive to decay chains in which only one chain produces a
photon, a broader set of models.

The Run I $\lgX$ search found good agreement with SM predictions in 86
$\invpb$ of data at a center-of-mass energy of 1.8 $\TeV$, except in
the $\lgmet$ channel, in which 16 events were observed with an
expectation of 7.6 $\pm$ 0.7, corresponding to a 2.7$\sigma$
excess. The Run I paper concluded: ``However, an excess of events with
0.7\% likelihood (equivalent to 2.7 standard deviations for a Gaussian
distribution) in one subsample among the five studied is an
interesting result, but it is not a compelling observation of new
physics. We look forward to more data in the upcoming run of the
Fermilab Tevatron.''~\cite{Jeff_PRL}.

Here we present the results from Run II with more than 10 times the
statistics of the Run I measurement.  We have repeated the $\lgX$
search with the same kinematic selection criteria in a data set
corresponding to an exposure of $\lumi\pm\dlumi$ $\invpb$, a higher
$\ppbar$ collision energy, 1.96 $\TeV$, and the CDF II
detector~\cite{CDFII}. The results correspond to the full data set
taken during the period March, 2002 through February, 2006, and
include data from the first third of this sample which have already
been presented~\cite{Loginov_all}. We give a detailed description of
the selection criteria, background calculations, and kinematic
distributions for the $\lgmet$ and $\llg$ channels. We also present
results for the first time for the $\emugX$ and $\lgg$ signatures.

This paper is organized as follows. Section~\ref{detector} gives a
brief description of the CDF II detector, emphasizing the changes from
Run I. Section~\ref{selection} presents the electron, muon, photon,
and $\met$ identification criteria, and the kinematic event selection
criteria. The data flow as additional selection criteria are added,
resulting in the measured number of events in each signature, is also
described. The standard model W and Z samples, used as control
samples, are described in Section~\ref{control}. Section~\ref{lgx}
gives an introduction to the selection of the Inclusive $\lgX$ event
sample. Section~\ref{lgmet} describes the selection of the $\lgmet$
signal sample, and presents the measured kinematic distributions.
Similarly, the $\llg$ signal sample selection and kinematic
distributions are presented in Section~\ref{llg}. A search for the
$\ell\gamma\gamma$ signature is briefly described in
Section~\ref{lgg}. Section~\ref{sm} summarizes the SM expectations
from $\Wg,\Wgg,\Zg,\Zgg$ production, and backgrounds from
misidentified photons, $\met$, and/or leptons. Sections~\ref{results}
and~\ref{conclusions} summarize the results and present the
conclusions, respectively.

\section{The CDF II Detector}
\label{detector}

The CDF II detector is a cylindrically symmetric spectrometer designed
to study $\ppbar$ collisions at the Fermilab Tevatron based on the
same solenoidal magnet and central calorimeters as the CDF I
detector~\cite{CDFI}. Because the analysis described here is intended
to repeat the Run I search as closely as possible, we note especially
the differences from the CDF I detector relevant to the detection of
leptons, photons, and $\met$. The tracking systems used to measure the
momenta of charged particles have been replaced with a central outer
tracker (COT) with smaller drift cells~\cite{COT}, and an enhanced
system of silicon strip detectors~\cite{SVX}. The calorimeters in the
regions~\cite{CDF_coo} with pseudorapidity $|\eta| \gt 1$ have been
replaced with a more compact scintillator-based design, retaining the
projective geometry~\cite{cal_upgrade}.  The coverage in $\varphi$ of
the central upgrade muon detector (CMP) and central extension muon
detector (CMX) systems~\cite{muon_systems} has been extended; the
central muon detector (CMU) system is unchanged.

\section{Selection of $\lgX$ Events}
\label{selection}

In order to make the present search statistically {\it a priori}, 
the identification of leptons and photons is essentially the same as
in the Run I search~\cite{Jeff_PRD}, with only minor technical changes
due to the differences in detector details between the upgraded
CDF II detector and CDF I. 

The scope and strategy of the Run I analysis were designed to reflect
the motivating principles.  Categories of photon-lepton events were
defined {\it a priori} in a way that characterized the different
possibilities for new physics.  For each category, the inclusive event
total and basic kinematic distributions can be compared with standard
model expectations.  The decay products of massive particles are
typically isolated from other particles, and possess large transverse
momentum and low rapidity.  The search is therefore limited to those
events with at least one isolated, central ($|\eta| < 1.0$) photon
with $\Et > 25 ~\GeV$, and at least one isolated, central electron or
muon with $\Et > 25 ~\GeV$.  These photon-lepton candidates are
further partitioned by angular separation.  Events where exactly one
photon and one lepton are detected nearly opposite in azimuth
($\Delta\varphi_{\ell\gamma} > 150^{\circ}$) are characteristic of a
two-particle final-state (two-body photon-lepton events), and the
remaining photon-lepton events are characteristic of three or more
particles in the final-state (multi-body photon-lepton events).  The
multi-body photon-lepton events are then further studied for the
presence of additional particles: photons, leptons, or the missing
transverse energy associated with weakly interacting neutral
particles.

In the subsections below we describe the real-time (``online'') event
selection criteria by the trigger system, and the subsequent event
selection ``offline'', including the selection of electrons, muons,
and photons, the rejection of jet background for leptons and photons
by track and calorimeter ``isolation'' requirements, and the
construction of the missing transverse energy $\met$ and total
transverse energy $\Ht$.

\subsection{The Online Selection by the Trigger System}
\label{trigger}

A 3-level trigger system~\cite{CDFII}  selects events 
with a high transverse momentum ($\Pt$)~\cite{EtPt} lepton ($\Pt >
18~\GeV$) or
photon ($\Et > 25~\GeV$) in the central region, $|\eta|
\lesssim 1.0$. The trigger system selects photon and electron
candidates from clusters of energy in the central electromagnetic
calorimeter. Electrons are distinguished from photons by the presence
of a COT track pointing at the cluster. The muon trigger requires a
COT track that extrapolates to a track segment (``stub'') in the muon
chambers~\cite{stub}. At each trigger level all transverse momenta are
calculated using the nominal center of the interaction region along
the beam-line, $z=0$~\cite{CDF_coo}.
%xxx Andrei we need to describe the triggers and subsequent samples actually
%used (I think this is two paragraphs).

%\subsection{Overview of Event Selection From Within the Recorded Data}
\subsection{Overview of Event Selection}
\label{offline}
\begin{figure}[!t]
\begin{center}
%\hspace*{-0.1in}
\includegraphics*[angle =0,width=0.48\textwidth]{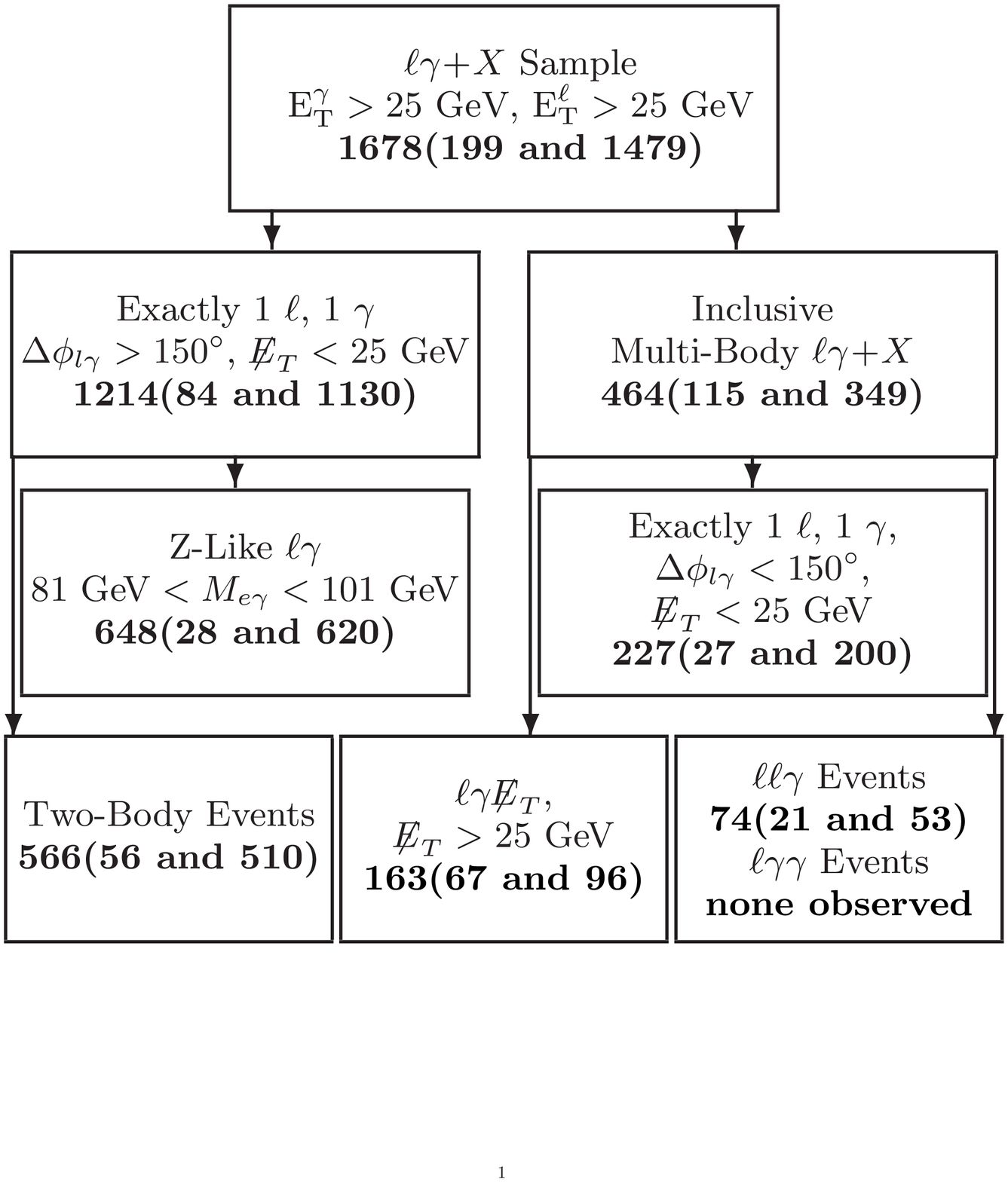}
\end{center}
\caption{$\lgX$ Sample: the subsets of inclusive lepton-photon events analyzed. The number of events in each subcategory is given as a sum of muons and electrons. The first term in parethesis refers to $\mugX$ while the latter refers to the $\egX$.}
\label{flowchart_lepton.figure}
\end{figure}

Inclusive $\lgal$ events (Fig.~\ref{flowchart_lepton.figure}) are
selected by requiring a central $\gamma$ candidate with $\Etgamma>25$
$\GeV$ and a central $e$ or $\mu$ with $\Etlepton>25$ $\GeV$
originating less than 60 cm along the beam-line from the detector
center and passing the ``tight'' criteria listed below.  All
transverse momenta, including that of the photon, are calculated using
the vertex within $\pm 5$ cm of the lepton origin that has the largest
scalar sum of transverse momentum from tracks associated to that
vertex. Both signal and control samples are drawn from this $\lgal$
sample (Fig.~\ref{flowchart_lepton.figure}).

%The $\lgal$ sample (Fig.~\ref{flowchart_lepton.figure}) is the sample
%from which both the signal and control samples are drawn.

Considering the control samples first, from the $\lgal$ sample we
select back-to-back events with exactly one photon and one lepton
(i.e. $\met<25~\GeV$); this is the dominant contribution to the
$\lgal$ sample, and has a large Drell-Yan component. A subset of this
sample is the `Z-like' sample, which provides the calibration for the
probability that an electron radiates and is detected as a photon, as
discussed in Section~\ref{fake_photons}. The remaining back-to-back
events are called the Two-Body Events and were not used in this
analysis.

All events which either have more than one lepton or photon, or in
which the lepton and photon are not back-to-back (and hence the event
cannot be a Two-Body event), are classified as `Inclusive Multi-Body
$\lgX$'. These are further subdivided into three categories: $\lgmet$
(Section~\ref{lgx}) (`Multi-Body $\lgmet$ Events'), for which the
$\met$ (Section~\ref{met}) is greater than 25 GeV , $\llg$
(Section~\ref{llg}) and $\lgg$ (Section~\ref{lgg}) (`Multi-Photon and
Multi-Lepton Events'), and events with exactly one lepton and exactly
one photon, which are not back-to-back. The events with exactly one
lepton and exactly one photon, which are not back-to-back were not
used in the analysis.
 
%Figure 1- the flow chart

\subsubsection{Electron Selection}
\label{electron}

An electron candidate passing the ``tight'' selection must have: a) a
high-quality track in the COT with $\Pt>0.5~\Et$, unless $\Et > 100$ $\GeV$, in
which case the $\Pt$ threshold is set to 25 $\GeV$; b) a good
transverse shower profile at shower maximum~\cite{cem} 
that matches the extrapolated track
position; c) a lateral sharing of energy in the two calorimeter towers
containing the electron shower consistent with that expected; and d)
minimal leakage into the hadron calorimeter~\cite{hadoem}.

Additional central electrons are required to have $\Et > 20~\GeV$ and
to satisfy the tight central electron criteria but with a track
requirement of only $\Pt>10$ $\GeV$ (rather than 0.5$\times\Et$), and
no requirement on a shower maximum measurement or lateral energy
sharing between calorimeter towers. Electrons in the end-plug
calorimeters ($1.2 < |\eta| < 2.0$) are required to have $\Et>
15~\GeV$, minimal leakage into the hadron calorimeter, a ``track''
containing at least 3 hits in the silicon tracking system, and a
shower transverse shape consistent with that expected, with a centroid
close to the extrapolated position of the
track~\cite{wenu_asymmetry_paper}.

\subsubsection{Muon Selection}
\label{muon}

A muon candidate passing the ``tight'' cuts must have: a) a
well-measured track in the COT with $\Pt > 25~\GeV$; b) energy
deposited in the calorimeter consistent with
expectations~\cite{muon_cal_cuts}; c) a muon ``stub''~\cite{stub} in
both the CMU and CMP, or in the CMX, consistent with the extrapolated
COT track~\cite{muon_stub_matching}; and d) COT timing consistent with
a track from a $\ppbar$ collision~\cite{muon_COT_timing}.

Additional muons are required to have $\Pt > 20~\GeV$ and to satisfy
the same criteria as for ``tight'' muons but with fewer hits required
on the track, or, alternatively, for muons outside the muon system
fiducial volume, a more stringent cut on track quality
but no requirement that there be a matching ``stub'' in the muon
systems~\cite{muon_track_quality}.

\subsubsection{Photon Selection}
\label{photon}

Photon candidates are required to have: no associated 
track with $\Pt>1$ $\GeV$;
at most one track with $\Pt<1$ $\GeV$, pointing at the calorimeter
cluster; good profiles in both transverse dimensions at shower
maximum; and minimal leakage into the hadron
calorimeter~\cite{hadoem}.

\begin{figure*}[!t]
\begin{center}
\hspace*{-0.1in}
\includegraphics*[angle =90,width=0.50\textwidth]{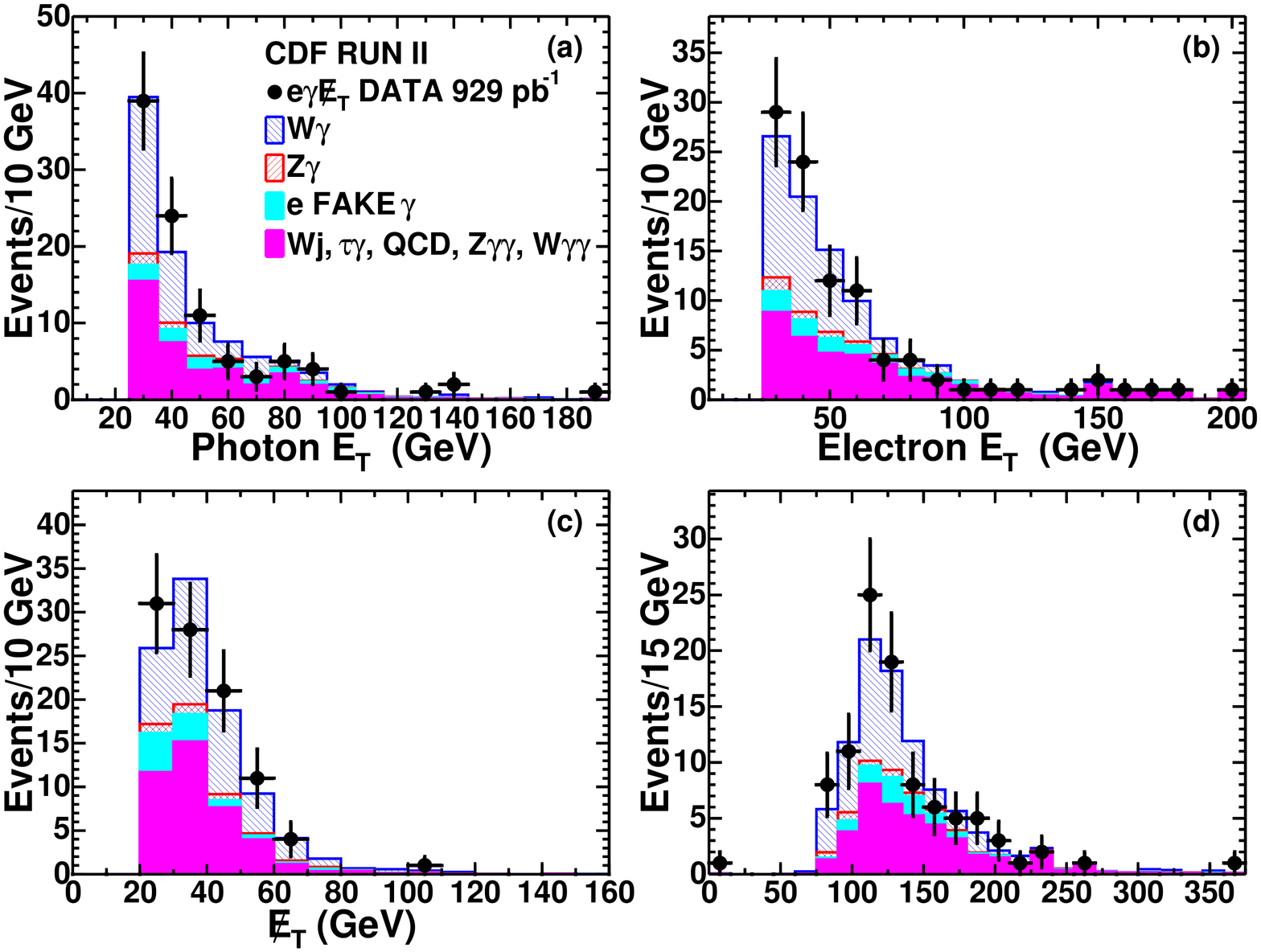}
\hfil
\includegraphics*[angle =90,width=0.50\textwidth]{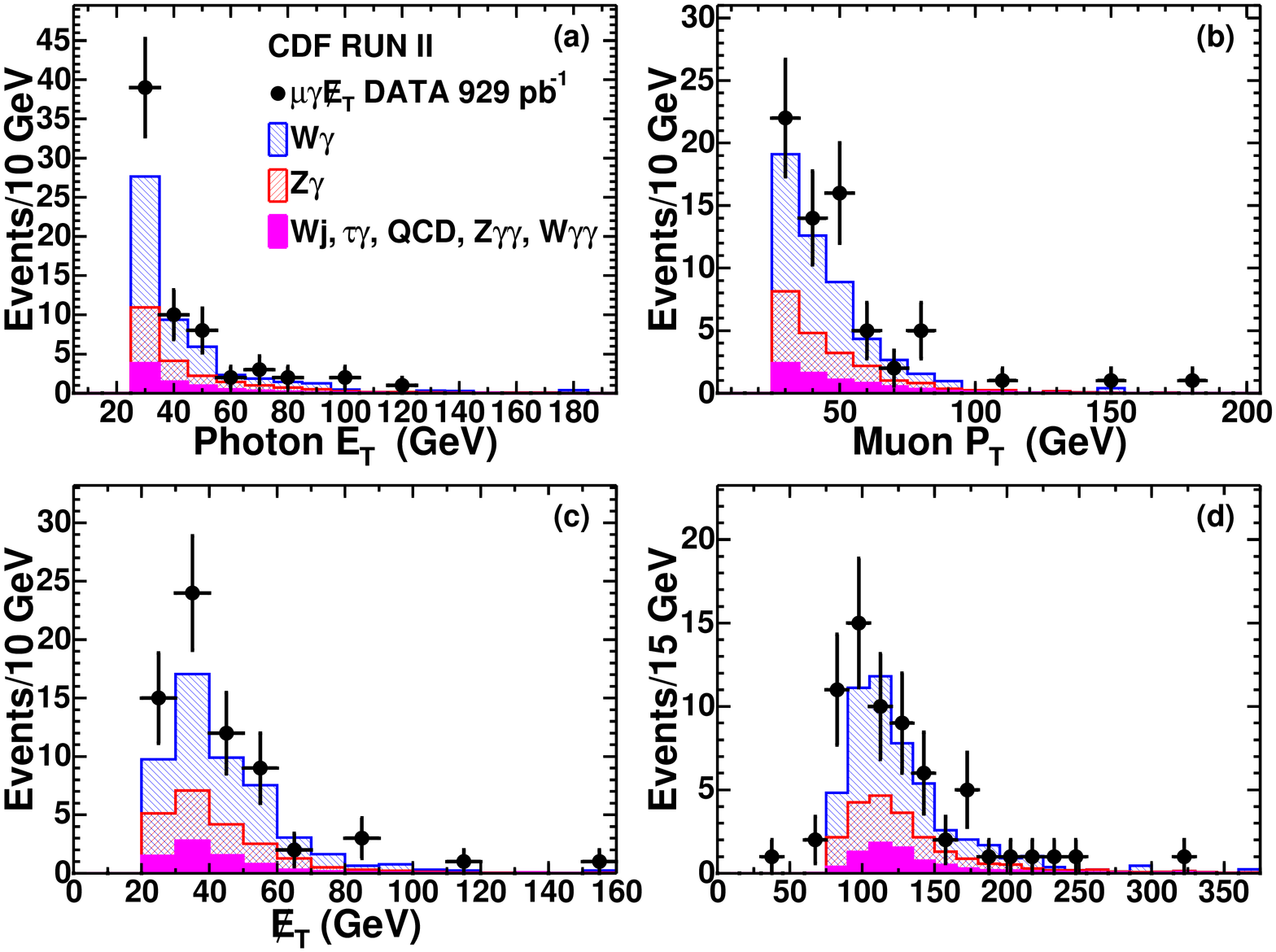}
\end{center}
\caption{ The distributions for events in the $\egmet$ sample (points
  in the left-hand four plots) and the $\mugmet$ sample (points in the
  right-hand four plots)
  for a) the $\Et$ of the photon; b) the $\Et$ of the lepton; c) the
  missing transverse energy, $\met$; and d) the transverse mass of the
  $\lgmet$ system.  The histograms show the expected SM contributions,
  including estimated backgrounds from misidentified photons and
  leptons.}
\label{wg_fig1_electrons}
\end{figure*}

\subsubsection{`Isolated' Leptons and Photons}
\label{iso}

To reduce background from photons or leptons from the decays of
hadrons produced in jets, both the photon and the lepton in each event
are required to be ``isolated''~\cite{isolation_nitpick}. The $\Et$
deposited in the calorimeter towers in a cone in $\eta-\varphi$
space~\cite{CDF_coo} of radius $R=0.4$ around the photon or lepton
position is summed, and the $\Et$ due to the photon or lepton is
subtracted. The remaining $\Et$ is required to be less than
$2.0~\GeV+0.02\times(\Et-20~\GeV)$ for a photon, or less than 10\% of
the $\Et$ for electrons or $\Pt$ for muons. In addition, for photons
the scalar sum of the $\Pt$ of all tracks in the cone must be less than
$2.0~\GeV+0.005\times\Et$.

% 2nd Figure for egmet and mugmet- double column- Ht et al
\begin{figure*}[!t]
\begin{center}
\hspace*{-0.1in}
\includegraphics*[angle =90,width=0.50\textwidth]{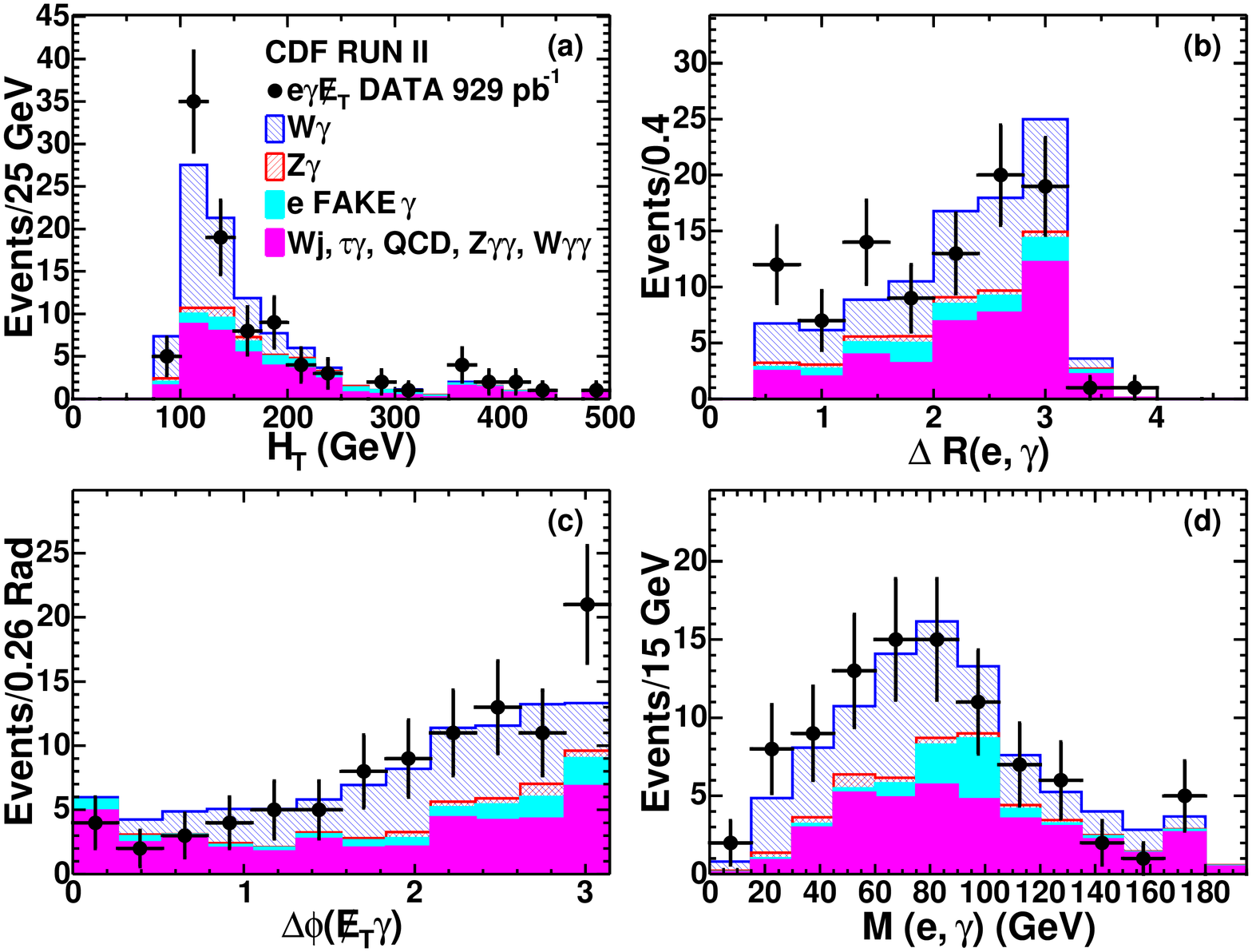}
\hfil
\includegraphics*[angle=90,width=0.50\textwidth]{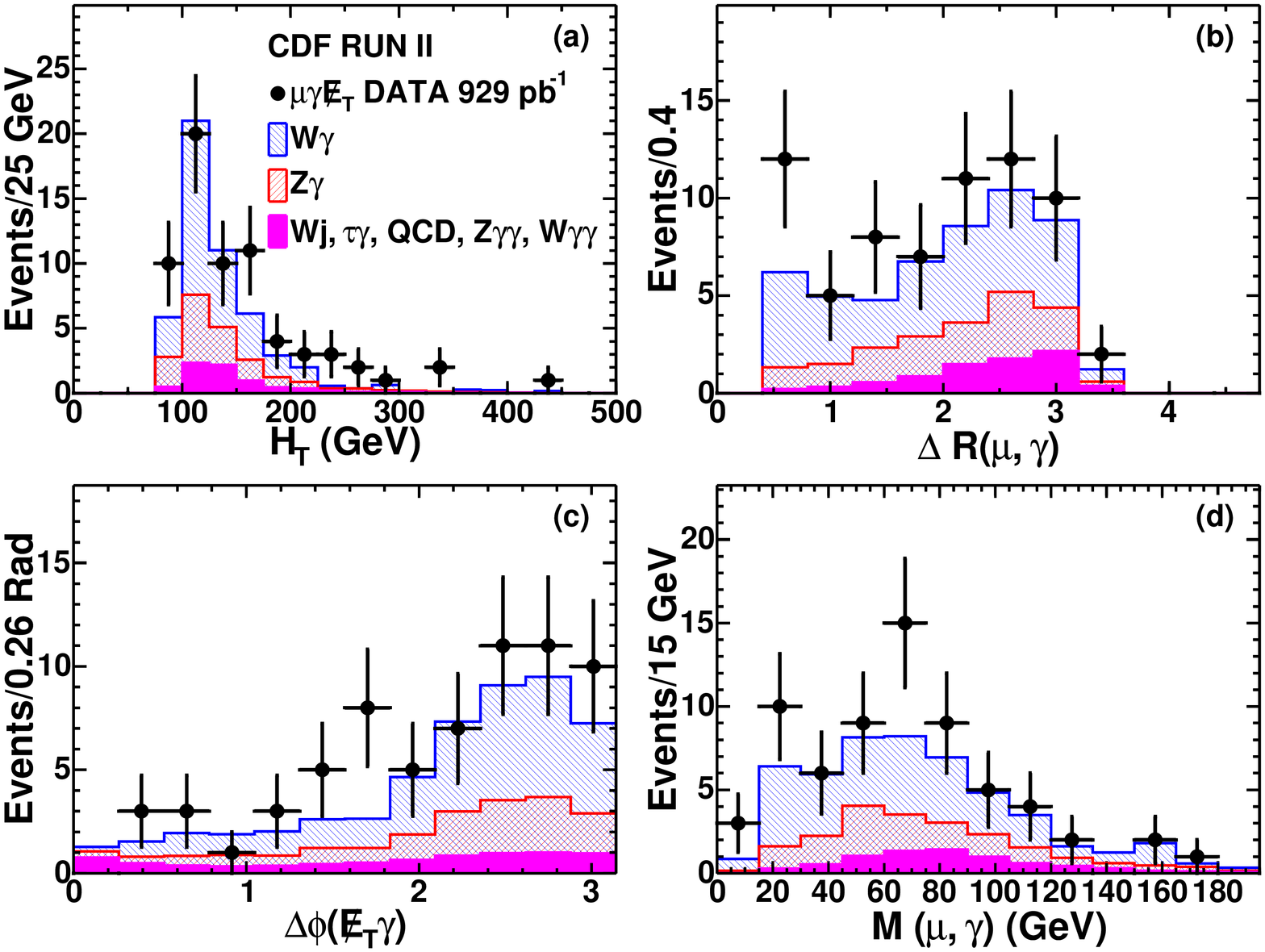}
\end{center}
\caption{ %The distributions for events in the $\egmet$ sample (points)
  The distributions for events in the $\egmet$ sample (points
  in the left-hand four plots) and the $\mugmet$ sample (points in the
  right-hand four plots)
  in a) $\Ht$, the sum of the transverse energies of the lepton,
  photon, jets and $\met$; b) the distance in $\eta$-$\phi$ space
  between the photon and lepton; c) the angular separation in $\phi$
  between the lepton and the missing transverse energy, $\met$; and
  d) the invariant mass of the $\lgal$ system.  The histograms show the
  expected SM contributions, including estimated backgrounds from
  misidentified photons and leptons.}
\label{wg_fig2_electrons}
\end{figure*}

\subsubsection{Missing Transverse Energy and $\Ht$}
\label{met}

Missing transverse energy $\met$ is calculated from the calorimeter
tower energies in the region $|\eta| < 3.6$. Corrections are then made
to the $\met$ for non-uniform calorimeter response~\cite{jet_corr} for
jets with uncorrected $\Et > 15$ $\GeV$ and $\eta < 2.0$, and for
muons with $\Pt > 20$ $\GeV$.

The variable $\Ht$ is defined for each event as the sum of the
transverse energies of the leptons, photons, jets, and $\met$ that
pass the above selection criteria.

\begin{figure}[!b]
\begin{center}
\hspace*{-0.1in}
\includegraphics*[angle =90,width=0.50\textwidth]{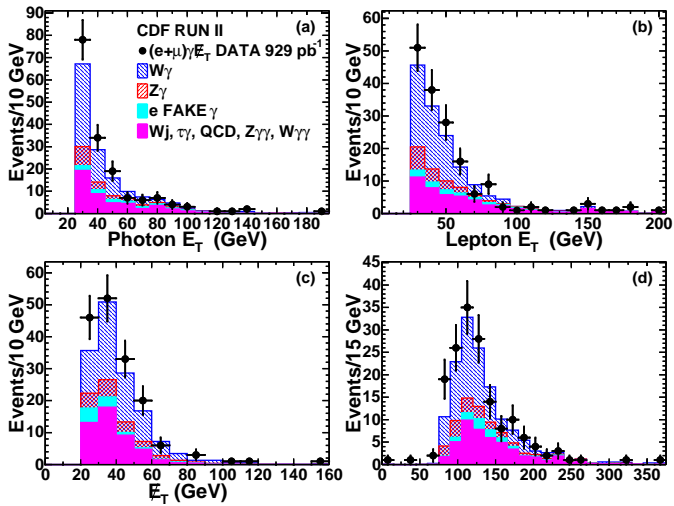}
\end{center}
\caption{ The distributions for events in the $\lgmet$ sample
  (points) in a) the $\Et$ of the photon; b) the $\Et$ of the lepton
  (e or $\mu$); c) the missing transverse energy, $\met$; and d) the
  transverse mass of the $\lgmet$ system.  The histograms show the
  expected SM contributions, including estimated backgrounds from
  misidentified photons and leptons.}
\label{wg_fig1_leptons}
\end{figure}

% 2nd Figure for lgmet; single column- Ht et al
\begin{figure}[!b]
\begin{center}
\hspace*{-0.1in}
\includegraphics*[angle =90,width=0.50\textwidth]{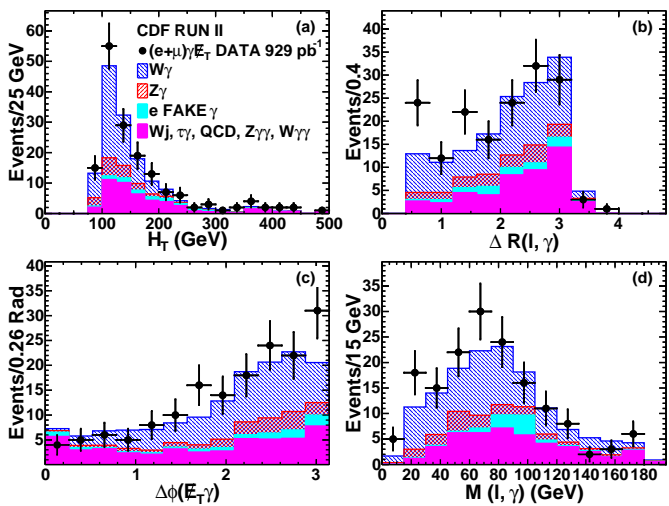}
\end{center}
\caption{ The distributions for events in the $\lgmet$ sample (points)
  in a) $\Ht$, the sum of the transverse energies of the lepton,
  photon, jets and $\met$; b) the distance in $\eta$-$\phi$ space
  between the photon and lepton; c) the angular separation in $\phi$
  between the lepton and $\met$; and d)
  the invariant mass of the $\lgal$ system. The histograms show the
  expected SM contributions, including estimated backgrounds from
  misidentified photons and leptons.}
\label{wg_fig2_leptons}
\end{figure}

\section{Control Samples}
\label{control}
Because we are looking for processes with small cross sections, and
hence small numbers of measured events, we use larger control samples
to validate our understanding of the detector performance and to
measure efficiencies and backgrounds. 

We use $W^\pm$ and $Z$ events reconstructed from the same inclusive
lepton datasets as control samples to ensure that the efficiencies for
high-$\Pt$ electrons and muons are well understood.  In addition, the
$W^\pm$ samples provide the control samples for the understanding of
$\met$. The selection criteria for the $W$ samples require a tight
lepton and $\met>25~\GeV$.  We find $\nofwenu$ $\Wenu$ events and
$\nofwmunu$ $\Wmunu$ events. For the $Z$ samples we require two
leptons, at least one of which satisfies the tight criteria.  We find
$\nofzee$ $\Zee$ events and $\nofzmumu$ $\Zmumu$ events.  The photon
control sample is constructed from $\Zee$ events in which one of the
electrons radiates a high-$\Et$ $\gamma$ such that the $\eg$ invariant
mass is within 10 $\GeV$ of the $Z$ mass.

\section{The Inclusive $\lgX$ Event Sample}
\label{lgx}

A total of $\noflg$ events, $\nofeg$ inclusive $\eg$ and $\nofmug$
inclusive $\mug$ candidates, pass the $\lgal$ selection criteria. Of
the $\nofeg$ inclusive $\eg$ events, $\nofegdphi$ have the electron
and photon within 30$\degs$ of back-to-back in $\varphi$, $\met< 25$
$\GeV$, and no additional leptons or photons. These are dominated by
$\Zee$ decays in which one of the electrons radiates a high-$\Et$
photon while traversing material before entering the COT active
volume, leading to the observation of an electron and a photon
approximately back-to-back in $\varphi$, with an $\eg$ invariant mass
close to the $Z$ mass.

%
% Figure 1 of llg- e and mu Et, etc. (way out of place, but that's Latex)
%
\begin{figure*}[!t]
\vspace*{-0.2in}
\begin{center}
\hspace*{-0.1in}
\includegraphics*[angle=90,width=0.50\textwidth]{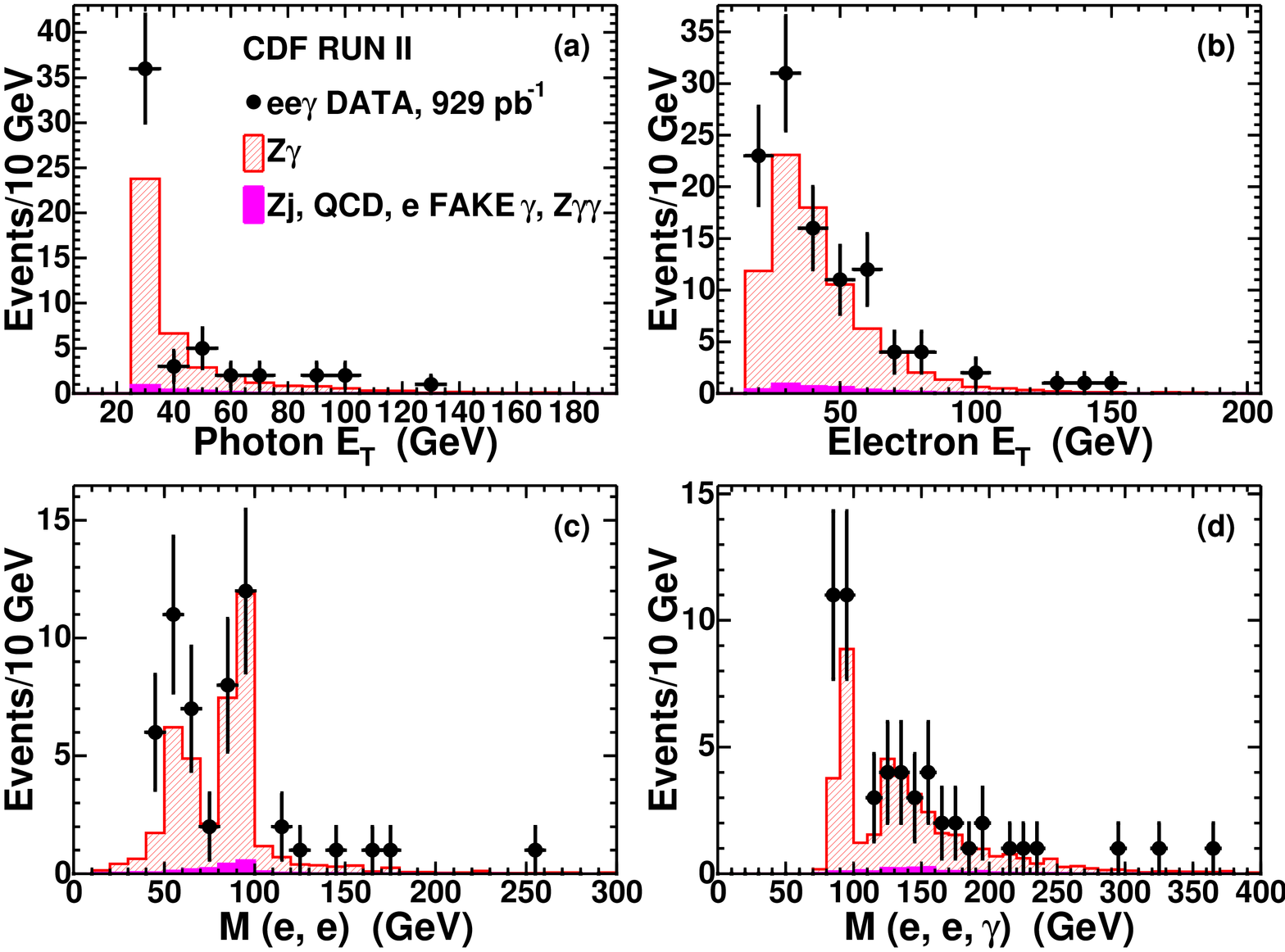}
\hfil
\includegraphics*[angle=90,width=0.50\textwidth]{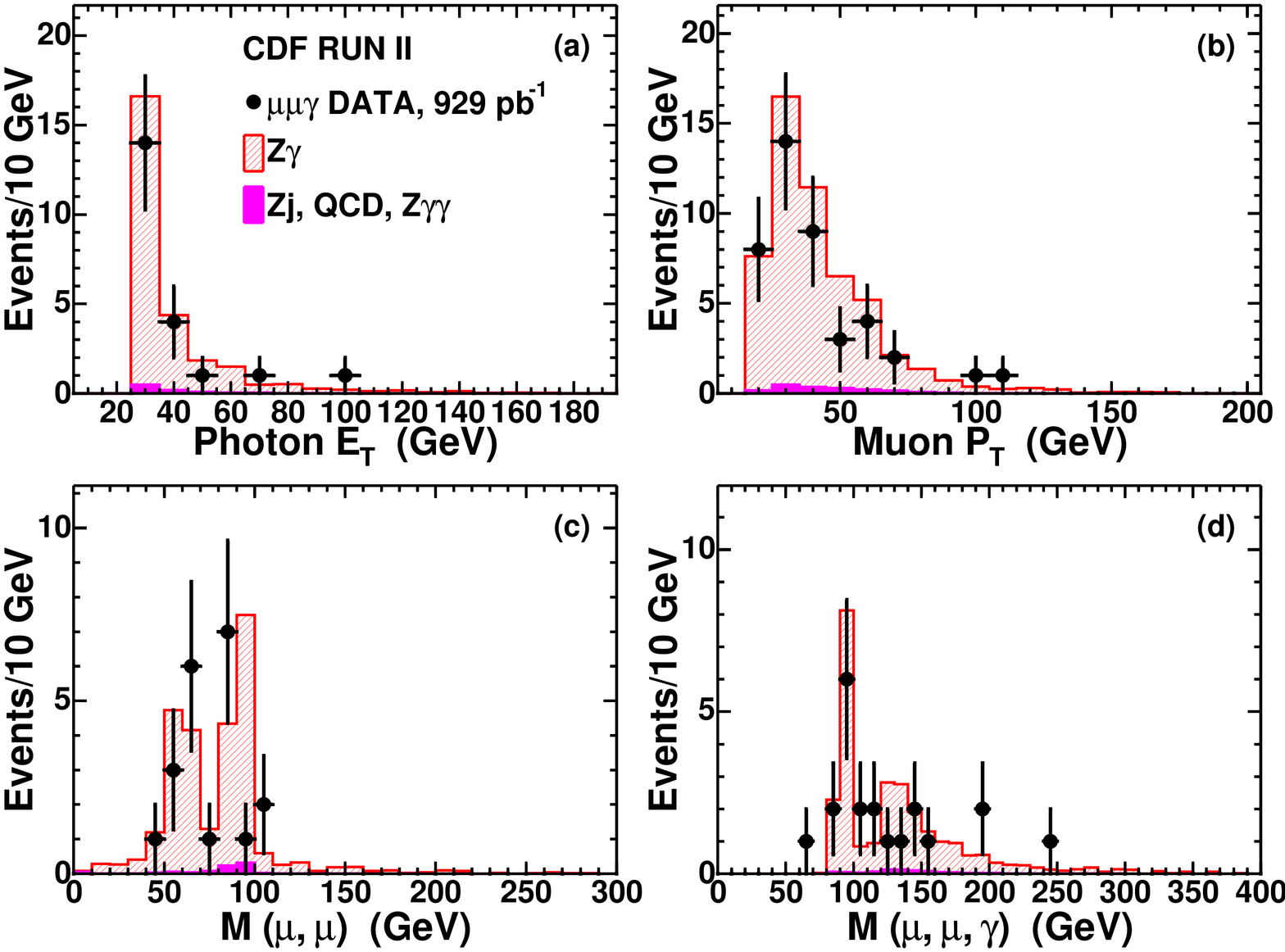}
\end{center}
\caption{ %The distributions for events in the $\eeg$ sample (points)
  The distributions for events in the $\eeg$ sample (points in the
  left-hand four plots) and the $\mumug$ sample (points in the
  right-hand four plots) in a) the $\Et$ of the photon; b) the $\Et$
  ($\Pt$) of the electrons (muons) (two entries per event); c) the
  2-body mass of the dilepton system; and d) the 3-body mass
  $M_{\llg}$. The histograms show the expected SM contributions.}
\label{zg_fig1_electrons}
\end{figure*}
%
% 2nd Figure 1 of llg- leptons Et, etc. (way out of place, but that's Latex)
%
\begin{figure}[!b]
\vspace*{-0.2in}
\begin{center}
\hspace*{-0.1in}
\includegraphics*[angle=90,width=0.50\textwidth]{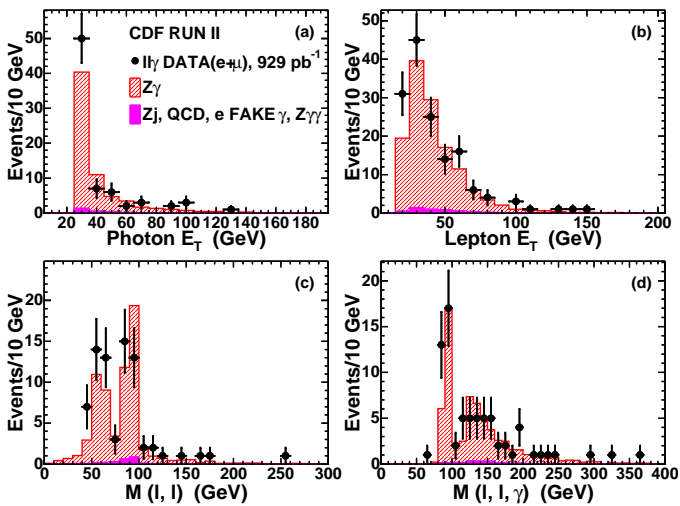}
\end{center}
\caption{ The distributions for events in the $\llg$ sample (points)
in a) the $\Et$ of the photon; b) the $\Et$ of the leptons (two
entries per event); c) the 2-body mass of the dilepton system; and d)
the 3-body mass $M_{\llg}$. The histograms show the expected SM
contributions.}
\label{zg_fig1_leptons}
\end{figure}

\section{The Inclusive $\lgmet$ Event Sample}
\label{lgmet}
The first search we perform is in the $\lgmet+X$ subsample, defined by
requiring that an event contain $\met> 25~\GeV$ in addition to the
$\gamma$ and ``tight'' lepton. Of the $\noflg$ $\lgal$ events,
$\nofeglgmet$ $e\gamma\met$ events and $\nofmuglgmet$ $\mu\gamma\met$
events pass the $\met$ requirement.

\subsection{Kinematic Distributions in the Electron and Muon Samples} 
The muon and electron signatures have different
backgrounds and detector resolutions, among other differences. While
these are corrected for, it is useful to plot the observed
distributions separately before combining them. We show both the
individual sample distributions as well as the final combined
plot~\cite{no_overflows}.

\subsubsection{Distributions in Photon  $\Et$, Lepton  $\Et$, $\met$, and 3-Body Transverse Mass}
Figure~\ref{wg_fig1_electrons} shows the observed distributions in a)
the $\Et$ of the photon; b) the $\Et$ of the lepton; c) $\met$; and d)
the transverse mass of the $\lgmet$ system, where $\rm{M_T} = [{(\rm
E_T^\ell+E_T^\gamma + \met)^2}$ - $(\lepvec + \phovec +
\metvec)^2]^{1/2}$. The left-hand set of four plots shows the
distributions for electrons; the right-hand set shows the
distributions for muons.

% Ht, Met etc. figures

\subsubsection{Distributions in $\Ht$,  $\Delta\phi_{\ell\gamma}$,$\Delta\phi_{\ell\met}$, $M_e\gamma$}

Figure~\ref{wg_fig2_electrons} shows the distributions for the
$\egmet$ sample (left) and $\mugmet$ sample (right) in a) $\Ht$, the
sum of the transverse energies of the lepton, photon, jets, and $\met$;
b) the distance in $\eta$-$\phi$ space between the photon and lepton;
c) the angular separation in $\phi$ between the lepton and the missing
transverse energy, $\met$; and d) the invariant mass of the $\lgal$
system.  The histograms show the expected SM contributions, including
estimated backgrounds from misidentified photons and leptons.

The electron and muon kinematic distributions are combined in
Fig.~\ref{wg_fig1_leptons} and Fig.~\ref{wg_fig2_leptons}. There
is very good agreement with the expected standard model shapes.

%%%%%%%%%%%%%%%%%%%%%%%%%%%%%%%%%%%%%%%%%%%%%%%%%%%%%%%%%%%%%%%%%%%%%%%%%%%%%%%
% 2nd set of figures for the llg search- dbl column Ht+Delta-R, single col
% lepton ditto, and e and mu met 

\begin{figure*}[!t]
\vspace*{-0.2in}
\begin{center}
\hspace*{-0.1in}
\includegraphics*[angle=90,width=0.50\textwidth]{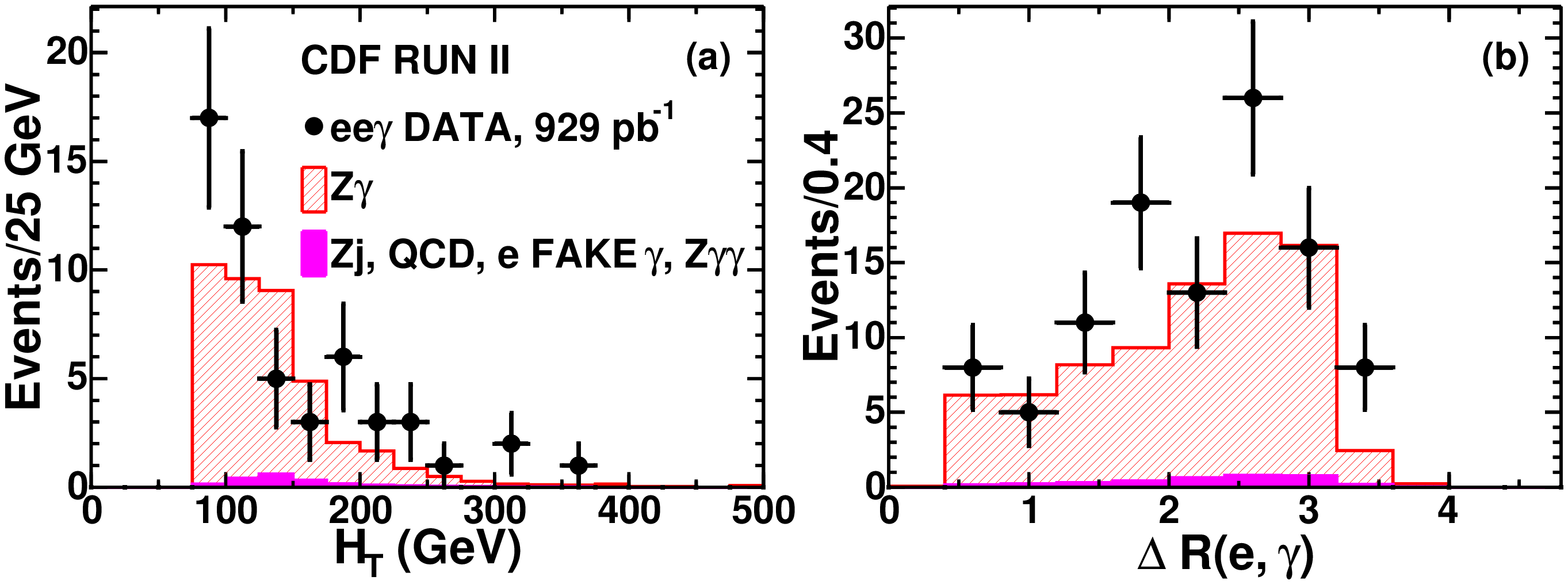}
\hfil
\includegraphics*[angle=90,width=0.50\textwidth]{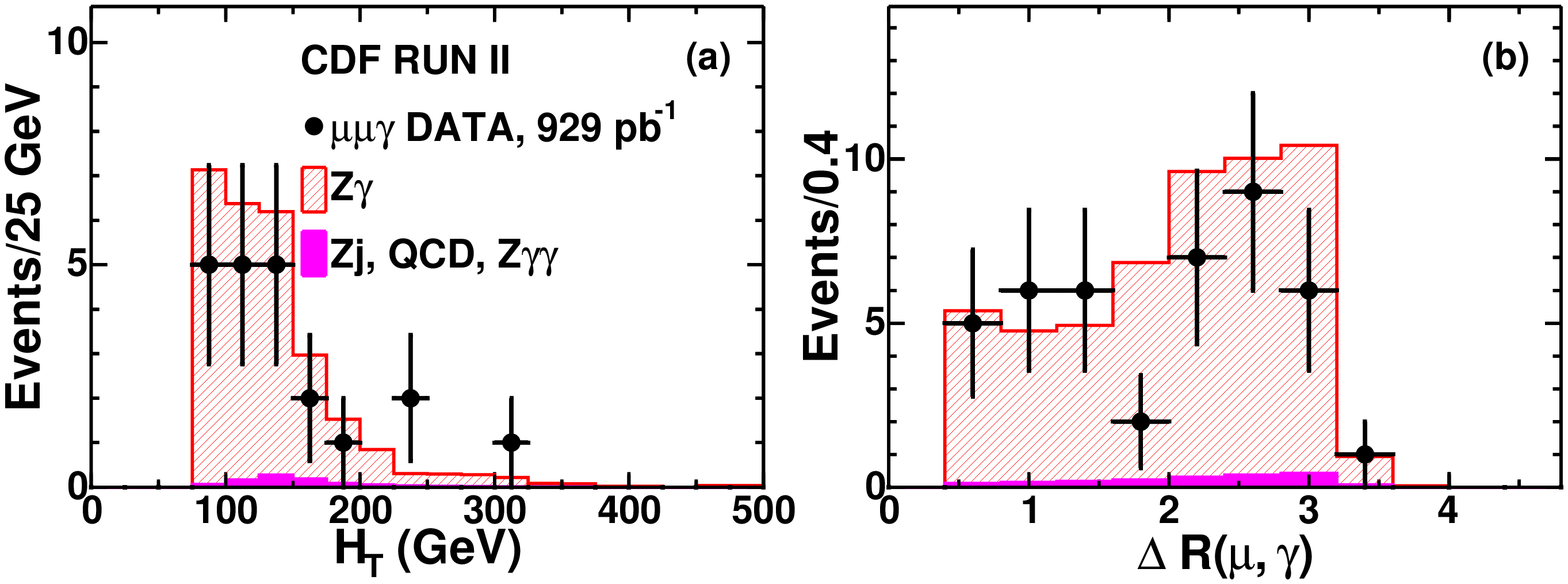}
\end{center}
\caption{
The distributions for events in the $\eeg$ sample (points in the
left-hand two plots) and the $\mumug$ sample (points in the right-hand
two plots) in a) $\Ht$, the sum of the transverse energies of the
lepton, photon, jets and $\met$; b) the distance in $\eta$-$\phi$
space between the photon and each of the two leptons (two entries
per event). The histograms show the expected SM contributions,
including estimated backgrounds from misidentified photons and
leptons.}
\label{zg_fig2_electrons}
\end{figure*}

\begin{figure}[!b]
\vspace*{-0.2in}
\begin{center}
\hspace*{-0.1in}
\includegraphics*[angle=90,width=0.50\textwidth]{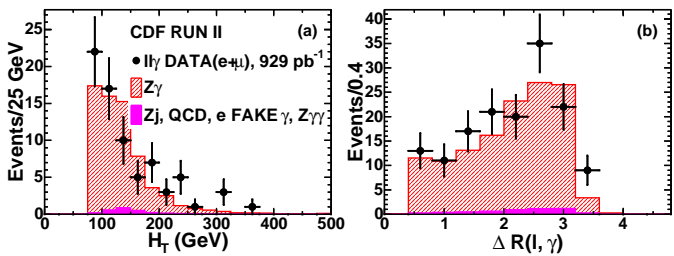}
\end{center}
\caption{ The distributions for events in the $\llg$ sample (points)
in a) $\Ht$, the sum of the transverse energies of the lepton, photon,
jets and $\met$; b) the distance in $\eta$-$\phi$ space between the
photon and each of the two leptons (two entries per event). The
histograms show the expected SM contributions, including estimated
backgrounds from misidentified photons and leptons.}
\label{zg_fig2_leptons}
\end{figure}

\begin{figure}[!b]
\vspace*{-0.2in}
\begin{center}
%\hspace*{-0.1in}
\includegraphics*[width=0.18\textwidth, angle=90,clip=]{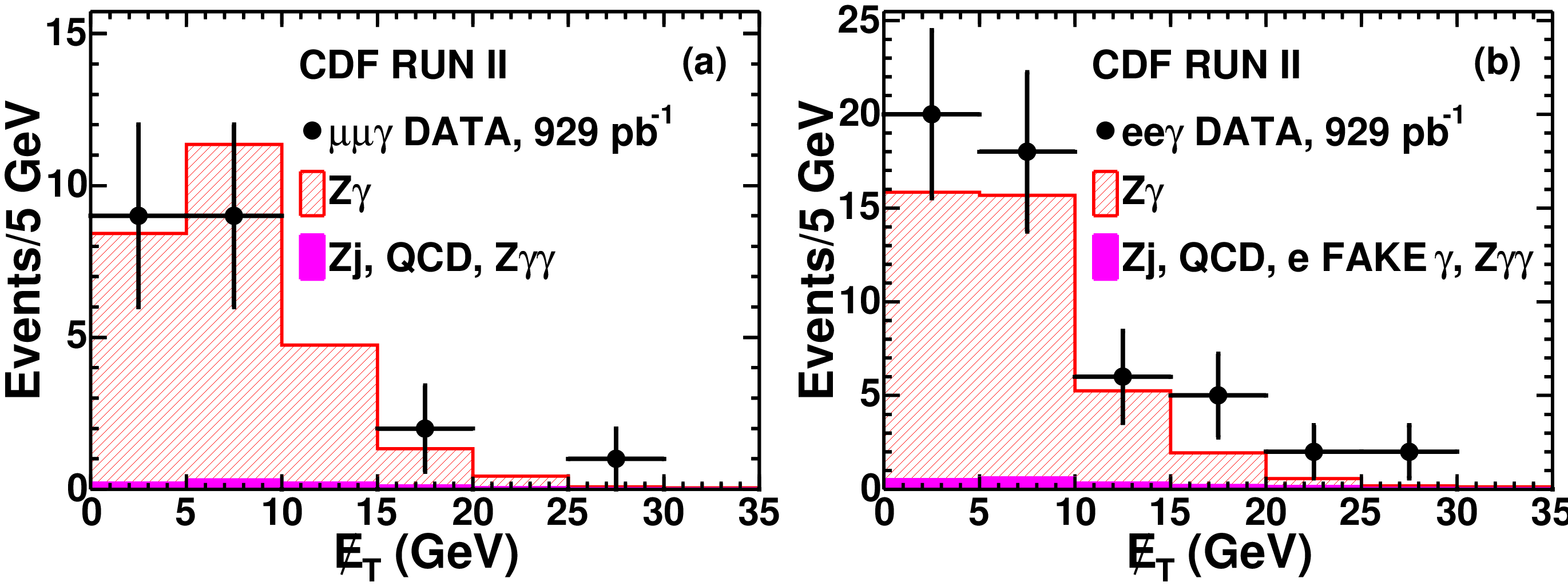}
\end{center}
\caption{The distributions in missing transverse energy $\met$
  observed in the inclusive search for a) $\mumug$ events and
  b) $\eeg$ events. The histograms show the expected SM
  contributions.}
\label{zg_fig3_leptons}
\end{figure}

\section{The Inclusive $\llg$ Event Sample}
\label{llg}

A second search, for the $\llg+X$ signature, is constructed by
requiring another $e$ or $\mu$ in addition to the ``tight'' lepton and
the $\gamma$.

The $\llg$ search criteria select $\noflgmultil$ events
($\nofegmultil$ $\eeg$ and $\nofmugmultil$ $\mumug$) of the $\noflg$
$\lgal$ events. No $e\mu\gamma$ events are observed.

\subsubsection{Distributions in Photon  $\Et$, Lepton  $\Et$, Dilepton
Invariant Mass, and $\llg$ Mass}

Figure~\ref{zg_fig1_electrons} shows the observed distributions in the
signature $\eeg$ (left-hand plots) and $\mumug$ channels (right-hand
plots) for: a) the $\Et$ of the photon; b) the $\Et$ of the electrons;
c) the 2-body mass of the dilepton system; and d) the 3-body mass
$M_{\eeg}$ or $M_{\mumug}$. For the $\Zg$ process occurring via
initial state radiation, the dilepton invariant mass $M_{\ell\ell}$
distribution is peaked around the $Z^0$-pole. For the final state
radiation, the three body invariant mass $M_{\llg}$ distribution is
peaked about the $Z^0$-pole.

The combined distributions for electrons and muons are shown in
Fig.~\ref{zg_fig1_leptons}.

\subsubsection{Distributions in $\Ht$ and $\Delta R_{\ell\gamma}$}

Figure~\ref{zg_fig2_electrons} shows the distributions for the $\eeg$
sample (left-hand plots) and $\mumug$ sample (right-hand plots) for:
a) $\Ht$, the sum of the transverse energies of the electron, photon,
jets and $\met$; b) and the distance in $\eta$-$\phi$ space between
the photon and each of the two leptons. The histograms show the
expected SM contributions, including estimated backgrounds from
misidentified photons and leptons.  The distributions for electrons
and muons are combined in Fig.~\ref{zg_fig2_leptons}.

\subsubsection{The Distributions in $\met$}

We do not expect SM events with large $\met$ in the $\llg$ sample; the
Run I $\eeggmet$ event was of special interest in the context of
supersymmetry~\cite{susy} due to the large value of $\met$ (55 $\pm$ 7
$\GeV$). Figure~\ref{zg_fig3_leptons} shows the distributions in
$\met$ for the $\mumug$ and $\eeg$ subsamples of the $\llg$ sample. We
observe 3 $\llg$ events with $\met > 25~\GeV$, compared to an
expectation of $0.6\pm0.1$ events.

\section{Search for the $\lgg$ Signature}
\label{lgg}

In some models of new phenomena the decay chain of each of a pair of
new heavy particles ends in a photon plus other
particles~\cite{susy}. One such signature that contains two photons
and is a subset of the $\lgX$ selection is $\lgg$.

The selection for the $\lgg$ search starts with a tight lepton and a
photon, each with $\Et > 25~\GeV$, from the same $\lgX$ sample as the
$\lgmet$ and $\llg$ searches. An additional photon with $\Et>25~\GeV$,
passing the same selection criteria as the first, is then required. We
observe no $\lgg$ events, compared to the expectation of
$\smnoflgmultig \pm \totdsysnoflgmultig$.

\section{Standard Model Expectations}
\label{sm}

\subsection{$\Wg$, $\Zg$, $\Wgg$, $\Zgg$}
\label{wgzgwggzgg}

The dominant SM source of $\lgal$ events is electroweak $W$ and
$\Zgstar$ production along with a $\gamma$ radiated from one of the
charged particles involved in the process~\cite{CDF_WZgamma}. The
number of such events is estimated using leading-order (LO) event
generators~\cite{MadGraph,Baur,CompHep}. Initial-state radiation is
simulated by the {\sc pythia} Monte Carlo (MC)
%simulated by the {\sc pythia} shower Monte Carlo (MC)
program~\cite{Pythia} tuned to reproduce the underlying event. The
generated particles are then passed through a full detector
simulation, and these events are then reconstructed with the same code
used for the data.

The expected contributions from $\Wg$ and $\Zgstar\plus\gamma$
production to the $\lgmet$ and $\llg$ searches are given in
Tables~\ref{lgmet.table} and~\ref{llg.table}, respectively. The
expected contributions to the $e\mug$ search are given in
Table~\ref{emug.table}. A correction for higher-order processes
(K-factor) that depends on both the dilepton mass and photon $\Et$ has
been applied~\cite{Baur_NLO}. In the $\lgmet$ signature we expect
$\wanoflglgmet \pm \totwadsysnoflglgmet$ events from $\Wg$ and
$\zanoflglgmet \pm \totzadsysnoflglgmet$ from $\Zgstar\plus\gamma$. In
the $\llg$ signature, we expect $\zanoflgmultil \pm
\totzadsysnoflgmultil$ events from $\Zgstar\plus\gamma$; the
contribution from $\Wg$ is negligible. The uncertainties on the SM
contributions include those from parton distribution functions (5\%),
factorization scale (2\%), K-factor (3\%), a comparison of
different MC generators ($\sim$ 5\%), and the luminosity (6\%).

We have used both {\sc madgraph}~\cite{MadGraph} and {\sc
comphep}\cite{CompHep} to simulate the triboson channels $\Wgg$ and
$Z\gamma\gamma$. The expected contributions are small,
$\waazaanoflglgmet \pm \waazaadsysnoflglgmet$ and $\waazaanoflgmultil
\pm \waazaadsysnoflgmultil$ events in the $\lgmet$ and $\llg$
signatures, respectively. The expected contributions from $\Wgg$ and
$\Zgstar\plus\gamma\gamma$ production to the $\lgg$ search are given
in Tables~\ref{lgmet.table} and~\ref{lgg.table}.

\subsection{Backgrounds from Misidentifications}
\label{fakes}

\subsubsection{``Fake'' Photons}
\label{fake_photons}

\begin{figure}[!t]
\centering
\includegraphics*[angle=90,width=0.48\textwidth]{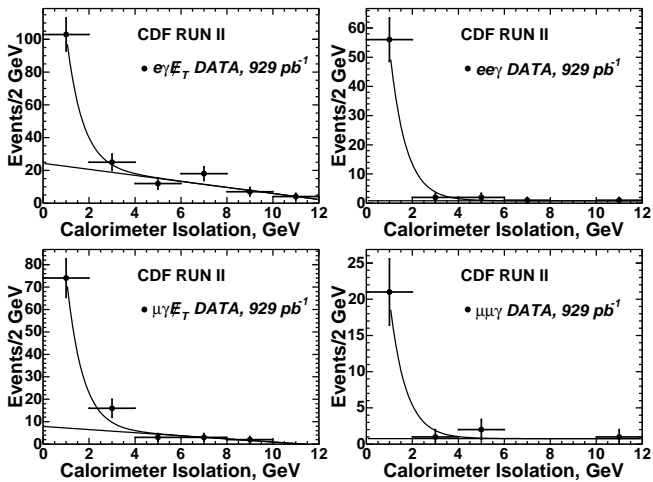}
\caption{The method and data used to estimate the number of background
events from jets misidentified as photons. For each of the four
samples, $\egmet$ (left top), $\eeg$ (right top), $\mugmet$ (left
bottom), and $\mumug$ (right bottom), the number of events is plotted
versus the total (electromagnetic plus hadronic) calorimeter energy,
$\Et^{Iso}$, in a cone in $\eta$-$\phi$ space around the photon. This
distribution is then fitted to the shape measured for electrons from
$\Zee$ decays plus a linear background.}
\label{wg_zg_jetfakes}
\end{figure}

\begin{figure}[!h]
\centering
\includegraphics*[angle=0,width=0.35\textwidth]{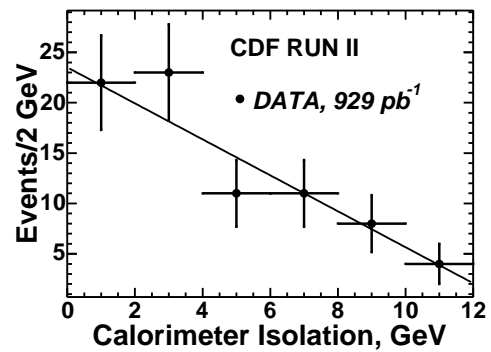}
\caption{The distribution in the total calorimeter
energy, $\Et^{Iso}$, in a cone in $\eta$-$\phi$ space around the fake
photon candidate. This distribution is then fitted with a linear
function.}
\label{pi0_isoet.figure}
\end{figure}

\begin{figure}[!t]
\centering
\includegraphics*[angle=0,width=0.35\textwidth]{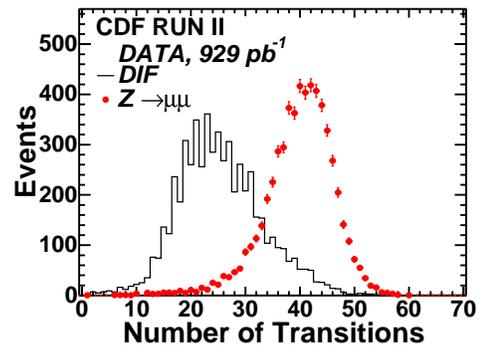}
\caption{The method and data used to estimate the number of background
muons from low-momentum hadrons decaying in flight. The number of
transitions in muons in the $\Zmumu$ sample is shown as points. The
number of transitions in muons in the sample enriched in hadron decays
is shown in the histogram, the so called decay-in-flight (``DIF'')
sample. The selection criteria for the DIF sample require a tight muon
with large impact parameter $d_0>$0.2 cm, at least one jet and
$\met>25~\GeV$.}
\label{ztight_mu_transitions_good_mu_transitions}
\end{figure}

High $\Pt$ photons are copiously created from hadron decays in jets
initiated by a scattered quark or gluon. In particular, mesons such as
the $\pi^0$ or $\eta$ decay to photons which may satisfy the photon
selection criteria. The numbers of lepton-plus-misidentified-jet
events expected in the $\lgmet$ and $\llg$ samples are determined by
measuring energy in the calorimeter nearby the photon candidate.

For each of the four samples, $\egmet$, $\mugmet$, $\eeg$, and
$\mumug$, Figure~\ref{wg_zg_jetfakes} shows the distribution in the
total (electromagnetic plus hadronic) calorimeter energy, $\Et^{Iso}$,
in a cone of radius $R=0.4$ in $\eta$-$\phi$ space around the photon
candidate. This distribution is then fitted to the shape measured for
electrons from $\Zee$ decays plus a linear background.

To verify the linear behavior of the background we select a sample of
``fake photons'' by requiring the photon candidate fail the cluster
profile criteria. In addition we do not apply the calorimeter and
track isolation requirements. The distribution in the total
calorimeter energy, $\Et^{Iso}$, in a cone of radius $R=0.4$ in
$\eta$-$\phi$ space around the fake photon candidate is shown in
Fig.~\ref{pi0_isoet.figure}.

The predicted number of events with jets misidentified as photons is
$\nofljglgmet \pm \totdsysnofljglgmet$ for the $\lgmet$ signature and
$\nofljgmultil \pmasym{\totdsysnofljgmultil}{0.0}$ for $\llg$.
%$\nofljgmultil \pm \totdsysnofljgmultil$ for $\llg$.

%\begin{table*}[p]
\begin{table*}[!t]
\begin{center}
\caption{A comparison of the numbers of events predicted by the
SM and the observations for the $\lgmet$ signature. The SM predictions
are dominated by $\Wg$ and $\Zg$
production~\cite{MadGraph,Baur,CompHep}. Other contributions come from
$\Wgg$ and $\Zgg$, leptonic $\tau$ decays, and misidentified leptons,
photons, or $\met$.}
\begin{tabular}{lccc}
\hline
\multicolumn{4}{c} {
{\bf Lepton+Photon+$\bf\met$ Events, $\lum=\bf\lumi$ $\bfinvpb$}}\\
\hline
{\bf SM Source} & $e\gamma\met$ & $\mu\gamma\met$ & $(e+\mu)\gamma\met$ \\
\hline%------------------------------------------------------------------------
$W^{\pm}\gamma$                         & %-------------
$\wanofeglgmet  \pm \totwadsysnofeglgmet$     &
$\wanofmuglgmet \pm \totwadsysnofmuglgmet$    &
$\wanoflglgmet  \pm \totwadsysnoflglgmet$     \\
$\Zgstar + \gamma$                    & %-------------
$\zanofeglgmet  \pm \totzadsysnofeglgmet$     &
$\zanofmuglgmet \pm \totzadsysnofmuglgmet$    &
$\zanoflglgmet  \pm \totzadsysnoflglgmet$     \\
$W^{\pm}\gamma\gamma$                   & %-------------
$\waanofeglgmet  \pm \totwaadsysnofeglgmet$   &
$\waanofmuglgmet \pm \totwaadsysnofmuglgmet$  &
$\waanoflglgmet  \pm \totwaadsysnoflglgmet$   \\
$\Zgstar + \gamma\gamma$              & %-------------
$\zaanofeglgmet  \pm \totzaadsysnofeglgmet$   &
$\zaanofmuglgmet \pm \totzaadsysnofmuglgmet$  &
$\zaanoflglgmet  \pm \totzaadsysnoflglgmet$   \\
$\ttg$                                  & %-------------
$\ttanofeglgmet  \pm \totttadsysnofeglgmet$   &
$\ttanofmuglgmet \pm \totttadsysnofmuglgmet$  &
$\ttanoflglgmet  \pm \totttadsysnoflglgmet$   \\
\hline%------------------------------------------------------------------------
%$\Z\goes \epem, e\goes\gamma$           & %-------------
$\ell e\met$, $e\goes\gamma$             & %-------------
$\efpnofeglgmet \pm \totefpdsysnofeglgmet$    &
$\efpnofmuglgmet \pm \totefpdsysnofmuglgmet$    &
$\efpnoflglgmet \pm \totefpdsysnoflglgmet$    \\
$W^{\pm}$+Jet faking $\gamma$                     & %-------------
%Jet faking $\gamma$                     & %-------------
$\nofejglgmet  \pm \totdsysnofejglgmet$       &
$\nofmujglgmet \pm \totdsysnofmujglgmet$      &
$\nofljglgmet  \pm \totdsysnofljglgmet$       \\
$W^{\pm}\gamma,\Zgstar\plus\gamma\goes\tau\gamma$     & %-------------
%$\tau\gamma$ contribution               & %-------------
$\wazataunofeglgmet  \pm \totwazataudsysnofeglgmet $  &
$\wazataunofmuglgmet \pm \totwazataudsysnofmuglgmet$  &
$\wazataunoflglgmet  \pm \totwazataudsysnoflglgmet $  \\
%QCD(Jets faking $\ell$ and $\met$)     & %-------------
QCD (Jets faking $\ell+\met$)                    & %-------------
$\jqcdnofeglgmet  \pm \totjqcddsysnofeglgmet $   &
$\jqcdnofmuglgmet \pmasym{\totjqcddsysnofmuglgmet}{0.0}$   &
%$\jqcdnofmuglgmet \pm \totjqcddsysnofmuglgmet$   &
$\jqcdnoflglgmet  \pm \totjqcddsysnoflglgmet $   \\
DIF (Decays-In-Flight)                         & %-------------
%$\difnofeglgmet  \pm \totdifdsysnofeglgmet $   &
--   &
$\difnofmuglgmet \pm \totdifdsysnofmuglgmet$   &
$\difnoflglgmet  \pm \totdifdsysnoflglgmet $   \\
\hline%\hline%-----------------------------------------------------------------
%%Total SM Prediction                    & 
%Total                                         & 
%$\smnofeglgmet  \pm \smdnofeglgmet (stat)$    &
%$\smnofmuglgmet \pm \smdnofmuglgmet(stat)$    &
%$\smnoflglgmet  \pm \smdnoflglgmet (stat)$    \\
%                                       & 
%$\pm \smdsysnofeglgmet (sys)$ &
%$\pm \smdsysnofmuglgmet(sys)$ &
%$\pm \smdsysnoflglgmet (sys)$ \\
%%\hline\hline%------------------------------------------------------------------
%\hline%------------------------------------------------------------------
%Total SM Prediction                               & 
{\bf Total SM  } &&&\\
{\bf Prediction}                            & 
%                                                 & 
{$\bf \smnofeglgmet  \pm \totdsysnofeglgmet $}    &
{$\bf \smnofmuglgmet \pm \totdsysnofmuglgmet$}    &
{$\bf \smnoflglgmet  \pm \totdsysnoflglgmet $}    \\
\hline%\hline%-----------------------------------------------------------------
%Observed in Data                       & 
{\bf Observed }&&&\\
{\bf in Data  }                                & 
{$\bf \nofeglgmet $}                         & 
{$\bf \nofmuglgmet$}                         & 
{$\bf \noflglgmet $}                         \\
\hline%------------------------------------------------------------------
\end{tabular}

\label{lgmet.table}
\end{center}
\end{table*}

For the $\lgg$ and $e\mug$ samples, due to the low statistics, the
above method cannot be used to find the numbers of background events
with a jet mis-identified as a photon. We instead measure the jet
$\Et$ spectrum in $\lgal+$jet, $\ell+$at~least two jets, and
$e\mu+$jet samples\cite{inclusive}, respectively, and then multiply by
the probability of a jet being misidentified as a photon,
$P^{jet}_{\gamma}(\Et)$, which is measured in data samples triggered
on jets. The uncertainty on the number of such events is calculated by
using the measured jet spectrum and the upper and lower bounds on the
$\Et$-dependent misidentification rate.

The misidentification rate is
$P^{jet}_{\gamma}=(6.5~\pm~3.3)\times10^{-4}$ for $\Et^{\gamma} = 25$
$\GeV$, and $(4.0~\pm~4.0)\times10^{-4}$ for $\Et^{\gamma} = 50$
$\GeV$ ~\cite{CDF_WZgamma}. The predicted number of events with jets
misidentified as photons is $\nofljgmultig \pm \totdsysnofljgmultig$
for the $\lgg$ signature and $\nofljgemp \pm \dnofljgemp$ for $e\mug$.

The probability that an electron undergoes hard bremsstrahlung and is
misidentified as a photon, $P^{e}_{\gamma}$, is measured from the
photon control sample.  The number of misidentified $\eg$ events
divided by twice the number of $ee$ events gives $P^{e}_{\gamma}$=(1.67
$\pm$ 0.07)\%. Applying this misidentification rate to electrons in
the inclusive lepton samples, we predict that $\efpnofeglgmet \pm
\totefpdsysnofeglgmet$ and $\efpnofegmultil \pm
\totefpdsysnofegmultil$ events pass the selection criteria for the
$\lgmet$ and $\llg$ searches, respectively. For the $\lgg$ search 
the estimated  background is $\efpnofegmultig \pm
\totefpdsysnofegmultig$ events.

\subsubsection{QCD Backgrounds to the $\lgmet$ and $\llg$ Signatures}
\label{qcd}

We have estimated the background due to events with jets misidentified
as $\lgmet$ or $\llg$ signatures by studying the total $\Pt$ of tracks
in a cone in $\eta-\varphi$ space of radius $R=0.4$ around the lepton
track. We estimate there are $\jqcdnoflglgmet \pm
\totjqcddsysnoflglgmet$ and $\jqcdnoflgmultil
\pmasym{\totjqcddsysnoflgmultil}{0.0}$ events in the $\lgmet$ and
$\llg$ signatures, respectively~\cite{QCD_background}.
%\totjqcddsysnoflglgmet$ and $\jqcdnoflgmultil \pm
%\totjqcddsysnoflgmultil$ events in the $\lgmet$ and $\llg$ signatures,
%respectively~\cite{QCD_background}.

There is a muon background that we expect escapes the above method.  A
low-momentum hadron, not in an energetic jet, can decay to a muon
forming a ``kink'' between the hadron and muon trajectories. In this
case a high-momentum track may be reconstructed from the initial track
segment due to the hadron and the secondary track segment from the
muon~\cite{decay_in_flight}. The contribution from this background is
estimated by identifying tracks consistent with a ``kink'' in the
COT. We count the number of times that, proceeding radially along a
COT track, a ``hit'' in the n+1 layer of sense-wires is on the other
side of the fitted track from the hit in the nth layer. Real tracks
will have hits distributed on both sides of the fit, and will
therefore have many ``transitions''. A mis-measured track from a
5-$\GeV$ $K^+$ (for example), on the other hand, will consist of two
intersecting low-momentum arcs fit by a high momentum track, and will
have a small number of transitions~\cite{Sasha}.

Figure~\ref{ztight_mu_transitions_good_mu_transitions} shows the
number of transitions in muons in the $\Zmumu$ control sample, and in
a sample enriched in hadron decays by selecting events with a large
$\met > 25~\GeV$, at least one jet and muon that have large impact
parameter $d_0>0.2~$cm. We estimate that there are $\difnofmuglgmet
\pm \totdifdsysnofmuglgmet$ and $\difnofmugmultil
\pmasym{\totdifdsysnofmugmultil}{0.0}$ events from decay-in-flight in
the $\mugmet$ and $\mumug$ samples, respectively.
%\totdifdsysnofmuglgmet$ and $\difnofmugmultil \pm
%\totdifdsysnofmugmultil$ events from decay-in-flight in the $\mugmet$
%and $\mumug$ samples, respectively.

\section{Results}
\label{results}

%\begin{table*}[p]
\begin{table*}[!t]
\begin{center}
\caption{A comparison of the numbers of events predicted by the
SM and the observations for the $\llg$ signature. The SM predictions
are dominated by $\Zg$ production~\cite{MadGraph,Baur,CompHep}. Other
contributions come from $\Zgg$, and misidentified leptons, photons, or
$\met$.}
\begin{tabular}{lccc}
\hline
\multicolumn{4}{c}{
{\bf Multi-Lepton + Photon Events, $\lum=\bf\lumi$ $\bfinvpb$}}\\ 
\hline
{\bf SM Source} & $\eeg$ & $\mumug$ & $(ee + \mu\mu)\gamma$  \\
\hline
$\Zgstar$            		          & 
$\zanofegmultil  \pm \totzadsysnofegmultil$     &
$\zanofmugmultil \pm \totzadsysnofmugmultil$    &
$\zanoflgmultil  \pm \totzadsysnoflgmultil$     \\
$\Zgstar+\gamma\gamma$ 	     	& 
$\zaanofegmultil  \pm \totzaadsysnofegmultil$   &
$\zaanofmugmultil \pm \totzaadsysnofmugmultil$  &
$\zaanoflgmultil  \pm \totzaadsysnoflgmultil$   \\
$W^{\pm}\gamma\gamma$                     & %-------------
$\waanofegmultil  \pm \waadnofegmultil$    &
$\waanofmugmultil \pmasym{\waadnofmugmultil}{0.0}$  &
%$\waanofmugmultil \pm \waadnofmugmultil$  &
$\waanoflgmultil  \pm \waadnoflgmultil$   \\
\hline
$\Zgstar+$Jet faking $\gamma$   & 
$\nofejgmultil  \pmasym{\totdsysnofejgmultil}{0.0}$       &
$\nofmujgmultil \pmasym{\totdsysnofmujgmultil}{0.0}$      &
$\nofljgmultil  \pmasym{\totdsysnofljgmultil}{0.0}$       \\
$\ell\ell e$, $e\goes\gamma$             & %-------------
$\efpnofegmultil \pm \totefpdsysnofegmultil$    &
$\efpnofmugmultil \pm \totefpdsysnofmugmultil$    &
$\efpnoflgmultil \pm \totefpdsysnoflgmultil$    \\
%QCD (Non-WZ) fakes                         & %-------------
QCD (Jets faking $\ell+\met$)                      & %-------------
$\jqcdnofegmultil  \pmasym{\totjqcddsysnofegmultil}{0.0} $   &
$\jqcdnofmugmultil \pmasym{\totjqcddsysnofmugmultil}{0.0}$   &
$\jqcdnoflgmultil  \pmasym{\totjqcddsysnoflgmultil}{0.0} $   \\
DIF (Decays-In-Flight)                     & %-------------
--                                                &
$\difnofmugmultil \pmasym{\totdifdsysnofmugmultil}{0.0}$   &
$\difnoflgmultil  \pmasym{\totdifdsysnoflgmultil}{0.0} $   \\
\hline%\hline
{\bf Total SM  }&&&\\
{\bf Prediction}                           & 
%Total SM Prediction             & 
%Total                                              & 
%                                                   & 
$\bf \smnofegmultil  \pm \totdsysnofegmultil $    &
$\bf \smnofmugmultil \pm \totdsysnofmugmultil$    &
$\bf \smnoflgmultil  \pm \totdsysnoflgmultil $    \\
\hline%\hline
{\bf Observed } &&&\\
{\bf in Data  }                                 & 
%Observed in Data                         & 
%Observed                                 & 
$\bf \nofegmultil$                           & 
$\bf \nofmugmultil$                          &                 
$\bf \noflgmultil$                           \\
\hline
\end{tabular}

\label{llg.table}
\end{center}
\end{table*}

The predicted and observed totals for the $\lgmet$ and $\llg$ searches
are shown in Tables~\ref{lgmet.table} and~\ref{llg.table},
respectively. We observe $\noflglgmet$ $\lgmet$ events, compared to the
expectation of $\smnoflglgmet \pm \totdsysnoflglgmet$ events. In the
$\llg$ channel, we observe $\noflgmultil$ events, compared to an
expectation of $\smnoflgmultil \pm \totdsysnoflgmultil$ events. There
is no significant excess in either signature.

The predicted and observed kinematic distributions for the $\egmet$
and $\mugmet$ signatures are compared in
Figs.~\ref{wg_fig1_electrons} and~\ref{wg_fig2_electrons}. The
corresponding distributions for the $\lgmet$ signature (the sum of
electrons and muons) are compared in Figs.~\ref{wg_fig1_leptons}
and~\ref{wg_fig2_leptons}.

\begin{table}[!b]
\begin{center}
\caption{A comparison of the numbers of events predicted by the
SM and the observations for the $\lgg$ signature.}
%The SM predictions are
%dominated by $\Zgg$ production~\cite{MadGraph,CompHep}. Dominant
%contribution comes from misidentified photons.}
\begin{tabular}{lccc}
\hline
\multicolumn{4}{c}{
{\bf Multi-Photon + Lepton Events, $\lum=\bf\lumi$ $\bfinvpb$}}\\ 
\hline
{\bf SM Source} & $\egg$ & $\mugg$ & $(e+\mu)\gamma\gamma$ \\
\hline
$W^{\pm}\gamma\gamma$                     & %-------------
$\waanofegmultig  \pm \waadnofegmultig$   &
$\waanofmugmultig \pm \waadnofmugmultig$  &
$\waanoflgmultig  \pm \waadnoflgmultig$   \\
$\Z\gamma\gamma$                        & %-------------
$\zaanofegmultig  \pm \zaadnofegmultig$   &
$\zaanofmugmultig \pm \zaadnofmugmultig$  &
$\zaanoflgmultig  \pm \zaadnoflgmultig$   \\
$\ell ee$, $\ell\gamma e$, $e\goes\gamma$ & %-------------
%$\Z\gamma$, $e\goes\gamma$                & %-------------
$\efpnofegmultig  \pm \totefpdsysnofegmultig$   &
$\efpnofmugmultig  \pmasym{\totefpdsysnofmugmultig}{0.0}$   &
%$\efpnofmugmultig  \pm \totefpdsysnofmugmultig$   &
$\efpnoflgmultig  \pm \totefpdsysnoflgmultig$   \\
$\ell jj$, $\ell\gamma j$, $j\goes\gamma$ & %-------------
$\nofejgmultig  \pm \totdsysnofejgmultig$       &
$\nofmujgmultig \pm \totdsysnofmujgmultig$      &
$\nofljgmultig  \pm \totdsysnofljgmultig$       \\
%\hline%------------------------------------------------------------------
%Total SM Prediction                       & 
%$\smnofegmultig  \pm \smdnofegmultig (stat)$     &
%$\smnofmugmultig \pm \smdnofmugmultig(stat)$    &
%$\smnoflgmultig  \pm \smdnoflgmultig (stat)$     \\
%                                          & 
%$\pm \smdsysnofegmultig (sys)$     &
%$\pm \smdsysnofmugmultig(sys)$    &
%$\pm \smdsysnoflgmultig (sys)$     \\
\hline%------------------------------------------------------------------
%Total SM Prediction                           & 
{\bf Total SM  } &&&\\
{\bf Prediction}                            & 
$\bf \smnofegmultig  \pm \totdsysnofegmultig $    &
$\bf \smnofmugmultig \pm \totdsysnofmugmultig$    &
$\bf \smnoflgmultig  \pm \totdsysnoflgmultig $    \\
\hline%------------------------------------------------------------------
%Observed in Data                         & 
{\bf Observed } &&&\\
{\bf in Data  }                                 & 
$\bf \nofegmultig$                           & 
$\bf \nofmugmultig$                          & 
$\bf \noflgmultig$                           \\
\hline%------------------------------------------------------------------
\end{tabular}

\label{lgg.table}
\end{center}
\end{table}

The predicted and observed kinematic distributions for the $\eeg$ and
$\mumug$ signatures are compared in Figs.~\ref{zg_fig1_electrons}
and~\ref{zg_fig2_electrons}.  The distributions for the $\llg$
signature are compared in
Figs.~\ref{zg_fig1_leptons},~\ref{zg_fig2_leptons}
and~\ref{zg_fig3_leptons}. We do find 3 $\llg$ events with $\met>
25~\GeV$, compared to an expectation of $0.6\pm0.1$ events,
corresponding to a likelihood of 2.4\%. We do not consider this
significant, and there is nothing in these 3 events to indicate they
are due to anything other than a fluctuation. We observe no $\llg$
events with multiple photons and so find no events like the $\eeggmet$
event of Run I.

The predicted and observed totals for the $\lgg$ and $e\mug$ searches
are shown in Tables~\ref{lgg.table} and~\ref{emug.table},
respectively. We observe no $\lgg$ or $e\mug$ events, compared to the
expectation of $\smnoflgmultig \pm \totdsysnoflgmultig$ and
$\smnoflgemug \pm \totdsysnoflgemug$ events, respectively.

\section{Conclusions}
\label{conclusions}

In Run I, in a sample of $\runonelumi$ $\invpb$ of $\ppbar$ collisions
at an energy of 1.8$~\TeV$, the CDF experiment observed a single clean
event consistent with having a pair of high-$\Et$ electrons, two
high-$\Et$ photons, and large $\met$~\cite{Toback_all}. A subsequent
search for ``cousins'' of the $\eeggmet$ signature in the inclusive
signature $\lgX$ found 16 events with a SM expectation of 7.6 $\pm$
0.7 events, corresponding in likelihood to a 2.7 $\sigma$
effect~\cite{Jeff_PRD,Jeff_PRL}.

To test whether something new was really there in either the
$\llggmet$ or $\lgmet$ signatures, we have repeated the $\lgX$ search
for inclusive lepton + photon production with the same kinematic
requirements as the Run I search, but with an exposure more than 10
times larger, $\lumi\pm\dlumi$ $\invpb$, a higher $\ppbar$ collision
energy, 1.96 $\TeV$, and the CDF II detector~\cite{CDFII}. Using the
same selection criteria makes this measurement an {\it a priori} test,
as opposed to the Run I measurement.
%We observe $\noflglgmet$ $\lgmet$ events, versus an expectation of
%$\smnoflglgmet \pm \totdsysnoflglgmet$ events from SM physics and
%background sources. In the $\llg$ channel, we observe $\noflgmultil$
%events, versus an expectation of $\smnoflgmultil \pm
%\totdsysnoflgmultil$ events. 
We find no significant excess in either
signature. We conclude that the 2.7 $\sigma$ effect observed in Run I
%measured with the same selection criteria,  
was a statistical fluctuation. 

With respect to the Run I $\eeggmet$ event, we observe no $\lgg$
events compared to an expectation of $\smnoflgmultig \pm
\totdsysnoflgmultig$ events.  The $\eeggmet$ event thus remains a
single event selected {\it a posteriori} as interesting, but whether
it was from SM $WW\gamma\gamma$ production, a rare background, or a
new physics process, we cannot determine.

\begin{table}[!t]
\begin{center}
\caption{A comparison of the numbers of events predicted by the
SM and the observations for the $e\mu\gamma$ signature. The SM
predictions are dominated by $\Zg$
production~\cite{MadGraph,Baur,CompHep}. Other contributions come from
$\Wg$, $\Zgg$, $\Wgg$, and misidentified leptons, photons, or $\met$.}
\begin{tabular}{lc}
\hline
\multicolumn{2}{c}{
{\bf $\bf e\mu$ + Photon Events, $\lum=\bf\lumi$ $\bfinvpb$}}\\
\hline
{\bf SM Source} & $e\mu\gamma+X$  \\
\hline
$\Zgstar + \gamma$            		          & 
$\zanoflgemp  \pm \zadnoflgemp$     \\
$W^\pm\gamma$            		          & 
$\wanoflgemp  \pmasym{\wadnoflgemp}{\wanoflgemp}$     \\
%$\wanoflgemp  \pm \wadnoflgemp$     \\
$\Z\gamma\gamma$ 	     	& 
$\zaanoflgemp  \pm \zaadnoflgemp$   \\
$W\gamma\gamma$ 	     	& 
$\waanoflgemp  \pm \waadnoflgemp$   \\
\hline
$e\mu j$, $j\rightarrow\gamma$   & 
$\nofljgemp  \pm \dnofljgemp$       \\
$ee\mu$, $e\goes\gamma$                & %-------------
%$to~be~filled$   \\
$\efpnofegemp  \pm \efpdnofegemp$   \\
\hline
%$\tau\gamma$ contribution               & %-------------
$W^{\pm}\gamma,\Zgstar\plus\gamma\goes\tau\gamma$     & %-------------
$\wazataunoflgemp  \pmasym{\wazataudnoflgemp}{\wazataunoflgemp} $  \\
%$\wazataunoflgemp  \pm \wazataudnoflgemp $  \\
\hline%\hline
%Total             & 
%$\smnoflgemug  \pm \smdnoflgemug    (stat)$    \\
%                                & 
%$             \pm \smdsysnoflgemug (sys) $    \\
%\hline
{\bf Total SM  } &\\
%{\bf Prediction} & $\bf \smnoflgemug \pmasym{\totdsysnoflgemug}{0.25}$    \\
{\bf Prediction} & $\bf \smnoflgemug \pm \totdsysnoflgemug$    \\
\hline%\hline
{\bf Observed } &\\
{\bf in Data  }                                 & 
$\bf \noflgemp$                              \\
\hline
\end{tabular}

\label{emug.table}
\end{center}
\end{table}

%Lastly, we observe no  $e\mug$ events, versus a SM expectation of 
%$\smnoflgemug \pm \totdsysnoflgemug$ events.

%\clearpage

%\input{lgx_runii_table_long}

\section{Acknowledgments}
\label{acknowledgments}

%\begin{acknowledgments}

We thank the Fermilab staff and the technical staffs of the
participating institutions for their vital contributions. Uli Baur,
Alexander Belyaev, Edward Boos, Lev Dudko, Tim Stelzer, and Steve
Mrenna were extraordinarily helpful with the SM predictions. This
work was supported by the U.S. Department of Energy and National
Science Foundation; the Italian Istituto Nazionale di Fisica Nucleare;
the Ministry of Education, Culture, Sports, Science and Technology of
Japan; the Natural Sciences and Engineering Research Council of
Canada; the National Science Council of the Republic of China; the
Swiss National Science Foundation; the A.P. Sloan Foundation; the
Bundesministerium f\"ur Bildung und Forschung, Germany; the Korean
Science and Engineering Foundation and the Korean Research Foundation;
the Particle Physics and Astronomy Research Council and the Royal
Society, UK; the Russian Foundation for Basic Research; the Comisi\'on
Interministerial de Ciencia y Tecnolog\'{\i}a, Spain; in part by the
European Community's Human Potential Programme under contract
HPRN-CT-2002-00292; and the Academy of Finland.
%\end{acknowledgments}

%====YOUAREHERE===============================\\
%\clearpage
%\newpage


\begin{thebibliography}{99}
\bibitem{SM} S.L.~Glashow,
Nucl. Phys. {\bf 22}, 588 (1961); S. Weinberg,
Phys. Rev. Lett. {\bf 19}, 1264 (1967);
A. Salam, Proc. 8th Nobel Symposium, Stockholm, (1979).
      
\bibitem{Toback_all}
F.~Abe \textit{et al.} (CDF Collaboration), Phys. Rev. D \textbf{59},
092002 (1999); F.~Abe \textit{et al.} (CDF Collaboration),
Phys. Rev. Lett. \textbf{81}, 1791 (1998); D.~Toback, Ph.D. thesis,
University of Chicago, 1997.

\bibitem{EtPt} Transverse momentum and energy are defined as $\Pt =
p\sin\theta$ and $\Et = E\sin\theta$, respectively.  
%
Missing $\rm E_T$ ($\metvec$) is defined by $\metvec = -\sum_{i} E_T^i
\hat{n}_i$, where i is the calorimeter tower number for $|\eta| <$ 3.6
(see Ref.~\cite{CDF_coo}), and $\hat{n}_i$ is a unit vector
perpendicular to the beam axis and pointing at the i$^{th}$
tower. We correct $\metvec$ for jets and muons. We define
the magnitude $\met=|\metvec|$.
%
We use the convention that ``momentum'' refers to $pc$ and ``mass'' to
$mc^2$.

%\cite{Allanach:2006fy}
%\bibitem{Allanach:2006fy}
\bibitem{lhc_wkgp} For a summary, see: 
  B.~C.~Allanach 	{\it et al.} (Les Houches working group),
  %``Les Houches 'Physics at TeV colliders 2005' Beyond the standard model
  %working group: Summary report,''
  arXiv:hep-ph/0602198.
  %%CITATION = HEP-PH 0602198;%%


\bibitem{susy_gauge} 
  D.~J.~H.~Chung, L.~L.~Everett, G.~L.~Kane, S.~F.~King, J.~D.~Lykken
  and L.~T.~Wang,
  %``The soft supersymmetry-breaking Lagrangian: Theory and applications,''
  Phys.\ Rept.\  {\bf 407}, 1 (2005)
  [arXiv:hep-ph/0312378]. The gravitino is very light, typically a few MeV.
  %%CITATION = HEP-PH 0312378;%%

\bibitem{susy_gravity}
  S.~P.~Martin,
  %``A supersymmetry primer,''
  arXiv:hep-ph/9709356.
  %%CITATION = HEP-PH 9709356;%%

\bibitem{kane_loop}
S.~Ambrosanio, G.~L.~Kane, G.~D.~Kribs, S.~P.~Martin, and S.~Mrenna,
Phys. Rev. Lett. \textbf{76}, 3498 (1996);
G.~L.~Kane and S.~Mrenna, Phys. Rev. Lett. \textbf{77}, 3502 (1996);
S.~Ambrosanio, G.~L.~Kane, G.~D.~Kribs, S.~P.~Martin, and S.~Mrenna, 
Phys. Rev. D \textbf{55}, 1372 (1997).


%\bibitem{gaugino_pair} 

%\bibitem{Tevatron_searches}
%  D.~Acosta {\it et al.}  [CDF Collaboration],
  %``Search for anomalous production of diphoton events with missing transverse
  %energy at {CDF} and limits on gauge-mediated supersymmetry-breaking models,''
%  Phys.\ Rev.\ D {\bf 71}, 031104 (2005)
%  [arXiv:hep-ex/0410053];
  %%CITATION = HEP-EX 0410053;%%

%  F.~Abe {\it et al.}  [CDF Collaboration],
%  %``Search for a technicolor omega(T) particle in events with a photon and a b
%  %quark jet at CDF,''
%  Phys.\ Rev.\ Lett.\  {\bf 83}, 3124 (1999)
%  [arXiv:hep-ex/9810031].
%  %%CITATION = HEP-EX 9810031;%%


%\bibitem{LEP_searches} 
%  G.~Abbiendi {\it et al.}  [OPAL Collaboration],
  %``Searches for gauge-mediated supersymmetry breaking topologies in e+ e-
  %collisions at LEP2,''
%  Eur.\ Phys.\ J.\ C {\bf 46}, 307 (2006)
%  [arXiv:hep-ex/0507048].
  %%CITATION = HEP-EX 0507048;%%

%\bibitem{HERA_searches} 
%\bibitem{Andreev:2003pm}
%  V.~Andreev {\it et al.}  [H1 Collaboration],
%   ``Isolated electrons and muons in events with missing transverse momentum at
  %HERA,''
%  Phys.\ Lett.\ B {\bf 561}, 241 (2003)
%  [arXiv:hep-ex/0301030].
  %%CITATION = HEP-EX 0301030;%%
%\cite{Chekanov:2003yt}
%\bibitem{Chekanov:2003yt}
%  S.~Chekanov {\it et al.}  [ZEUS Collaboration],
  %``Search for single-top production in e p collisions at HERA,''
%  Phys.\ Lett.\ B {\bf 559}, 153 (2003)
%  [arXiv:hep-ex/0302010].
  %%CITATION = HEP-EX 0302010;%%

\bibitem{LED}
 N.~Arkani-Hamed, S.~Dimopoulos and G.~R.~Dvali,
Phys.\ Lett.\ B {\bf 429}, 263 (1998)
  [arXiv:hep-ph/9803315].
  %%CITATION = HEP-PH 9803315;%%


\bibitem{excited_electron} 
D. Acosta \textit{et al.} (CDF Collaboration), Phys. Rev. Lett. \textbf{94},
101802 (2005). A similar search for an excited muon state is given in:
A.~Abulencia {\it et al.}  (CDF Collaboration),
Phys.\ Rev.\ Lett.\  {\bf 97}, 191802 (2006)
[arXiv:hep-ex/0606043].

\bibitem{Geraldine} See, for example,  
K.~Agashe and G.~Servant, JCAP {\bf 0502}, 002 (2005)
[arXiv:hep-ph/0411254].

\bibitem{Jeff_PRD} 
D. Acosta \textit{et al.} (CDF Collaboration), Phys. Rev. D
\textbf{66}, 012004 (2002); hep-ex/0110015.

\bibitem{Jeff_PRL} 
D. Acosta \textit{et al.} (CDF Collaboration), Phys. Rev. Lett. \textbf{89},
041802 (2002); hep-ex/0202044.

\bibitem{Jeff_thesis}
J.~Berryhill, Ph.D. thesis, University of Chicago, 2000.


\bibitem{CDFII} 
D. Acosta \textit{et al.} (CDF Collaboration), Phys. Rev. D
\textbf{71}, 032001 (2005).

%\bibitem{CDFII} 
%R. Blair \textit{et al.} (CDF Collaboration), CDF-II Technical Design
%Report, FERMILAB-PUB-96/390-E (1996).

\bibitem{Loginov_all} 
A.~Abulencia \textit{et al.} (CDF Collaboration), Phys. Rev. Lett. \textbf{97},
031801 (2006), hep-ex/0605097;
A.~Loginov for the CDF Collaboration, Eur. Phys. J. C \textbf{46},
s2.21-s2.31 (2006), hep-ex/0604036;
A. Loginov, Ph.D thesis, Institute for
Theoretical and Experimental Physics, Moscow, Russia, September,
2006, FERMILAB-THESIS-2006-48, hep-ex/0703011.

%\bibitem{Loginov_PRL} 
%A.~Abulencia \textit{et al.} (CDF Collaboration), Phys. Rev. Lett. \textbf{97},
%031801 (2006); hep-ex/0605097.
%
%\bibitem{Loginov_EPJC} 
%A.~Loginov for the CDF Collaboration, Eur. Phys. J. C \textbf{46},
%s2.21-s2.31 (2006); hep-ex/0604036.

\bibitem{CDFI} 
F. Abe \textit{et al.} (CDF Collaboration),
Nucl. Instrum. Methods A \textbf{271}, 387 (1988).

\bibitem{COT} 
A. Affolder \textit{et al.}, 
Nucl. Instrum. Methods A \textbf{526}, 249 (2004).

\bibitem{SVX} 
A. Sill \textit{et al.}, 
Nucl. Instrum. Methods A \textbf{447}, 1 (2000);
A. Affolder \textit{et al.},
Nucl. Instrum. Methods A \textbf{453}, 84 (2000); 
C.S. Hill,
Nucl. Instrum. Methods A \textbf{530}, 1 (2000).

\bibitem{CDF_coo} The CDF coordinate system of 
$r$, $\varphi$, and $z$ is cylindrical, with the $z$-axis along the
proton beam. The pseudorapidity is $\eta = -\ln(\tan(\theta/2))$.

\bibitem{cal_upgrade} S. Kuhlmann \textit{et al.}, 
Nucl. Instrum. Methods A \textbf{518}, 39 (2004).

\bibitem{muon_systems} 
The CMU system consists of gas proportional chambers in the region
$|\eta|<0.6$; the CMP system consists of chambers after an additional
meter of steel, also for $|\eta|<0.6$. The CMX chambers cover
$0.6<|\eta|<1.0$.

\bibitem{stub} A. V. Varganov, Ph.D. Thesis, University of Michigan, 2004; AAT
3137951.

\bibitem{cem} 
L.~Balka \textit{et al.}, Nucl. Instrum. Methods A {\bf 267}, 272
(1988).

\bibitem{hadoem} 
The fraction of electromagnetic energy allowed to leak into the hadron
compartment $\rm E_{had}/E_{em}$ must be less than $\rm
0.055\plus0.00045\times E_{em}(\GeV)$ for central electrons, less than
0.05 for electrons in the end-plug calorimeters, less than max[0.125,
$\rm 0.055\plus0.00045\times E_{em}(\GeV)$] for photons.

\bibitem{wenu_asymmetry_paper} 
D.~Acosta \textit{et al.} (CDF Collaboration),
Phys. Rev. D \textbf{71}, 051104 (2005); hep-ex/0501023.

\bibitem{muon_cal_cuts} 
The energy deposited in the calorimeter tower traversed by the muon
must be less than $2+max(0,0.0115\times(p-100))~\GeV $ in the
electromagnetic compartment and less than
$6+max(0,0.028\times(p-100))~\GeV$ in the hadronic compartment.

\bibitem{muon_stub_matching} 
The muon `stub' in the muon systems must be within 3, 5, and 6 cm of
the extrapolated COT track position, in the CMU, CMP, and CMX muon
systems, respectively.

\bibitem{muon_COT_timing}
A. Kotwal, H. Gerberich, and C. Hays, Nucl. Instrum. Methods A
{\bf 480}, 110 (2003).

\bibitem{muon_track_quality} 
For tight muons and tight electrons at least 5 hits in each of 3 axial
and 3 stereo layers of the COT are required; for loose muons with a
matching muon stub this is relaxed to 3 axial and 2 stereo.  Loose
muons without a matching stub have an additional cut on the $\chi ^2$
for the fit to the track.

\bibitem{isolation_nitpick} 
Note that this requirement is not a cut on the intrinsic properties of
the lepton or photon, but is instead a topological discriminant
between those physics processes producing leptons not close to jets
(signal) and those with leptons inside jets (presumably background).

%for photons: $0.055+0.00045E_{em}$ or 0.125
%for plug: 0.05
%for electrons: $0.055+0.00045E_{em}$

\bibitem{jet_corr} 
A. Bhatti \textit{et al.}, 
Nucl. Instrum. Methods A \textbf{566}, 375 (2006);
%submitted to Nucl. Instrum. Methods, %Oct. 2005; 
hep-ex/0510047.

\bibitem{no_overflows} 
There are no overflows in any of the distributions shown in the 
figures in this paper.

\bibitem{susy} 
S.~Ambrosanio, G.L.~Kane, G.D.~Kribs, S.P.~Martin, and S.~Mrenna,
Phys. Rev. D \textbf{55}, 1372 (1997); B.C.~Allanach, S.~Lola,
K.~Sridhar, Phys. Rev. Lett. \textbf{89}, 011801 (2002).

\bibitem{CDF_WZgamma} %CDF has measured $\Wg$  and $\Zg$  production 
D. Acosta \textit{et al.} (CDF Collaboration), Phys. Rev. Lett. \textbf{94},
041803 (2005).

\bibitem{MadGraph} 
T. Stelzer and W. F. Long, Comput. Phys. Commun. \textbf{81}, 357
(1994); F. Maltoni and T. Stelzer, J. High Energy Phys. \textbf{302},
27 (2003).

\bibitem{Baur} 
U. Baur, T. Han, and J. Ohnemus,
Phys. Rev. D \textbf{48}, 5140 (1993);
J. Ohnemus, Phys. Rev. D \textbf{47}, 940 (1993).

\bibitem{CompHep} 
E. Boos \textit{et al.} (The {\sc comphep} Collaboration),
Nucl. Instrum. Methods A \textbf{534}, 250, (2004); hep-ph/0403113.

\bibitem{Pythia} 
T.~Sjostrand, Comput. Phys. Commun. \textbf{82} (1994) 74;
S.~Mrenna, Comput. Phys. Commun. \textbf{101} (1997) 232.
 
\bibitem{Baur_NLO} 
U.~Baur, T.~Han and J.~Ohnemus, 
Phys.\ Rev.\ D {\bf 48}, 5140 (1993);
U.~Baur, T.~Han and J.~Ohnemus, 
Phys.\ Rev.\ D {\bf 57}, 2823 (1998); hep-ph/9710416; and U. Baur,
private communication. 
The K-factor $K_W$ applied to $\Wg$ MadGraph MC samples is $K_W
= 1.36$ for $M_W \le 76~\GeV$ and $K_W = 1.62 + 10^{-5}\times
P_T^\gamma - 0.386\times e^{-0.1\times P_T^\gamma}$ for $M_W \gt
76~\GeV$.  The K-factor $K_Z$ applied to $\Zg$ MadGraph MC samples is
$K_Z = 1.36$ for $M_Z \le 86~\GeV$ and $K_Z = 1.46 - 0.000728\times
P_T^\gamma- 0.125\times e^{-0.0615\times P_T^\gamma}$ for $M_Z \gt
86~\GeV$.
%
%See also Ref.~\cite{loginov_thesis}.
%Both the $\Wg$ and $Z\gamma$ K-factors are fixed at 1.36 for generated
%$\ell\nu$ masses below 76 $\GeV$ and for generated $\lplm$ masses
%below 86 $\GeV$. Above the poles the K-factors grow with
%$\Et^{\gamma}$ to be 1.62 and 1.53 at $\Et^{\gamma}=100$ $\GeV$ for
%$\Wg$ and $Z\gamma$, respectively. 

\bibitem{inclusive} Following the convention used throughout this
paper, these samples are inclusive and are defined by the minimum 
set of objects required. Any number of additional jets, leptons, or
photons, and any value of $\met$,  may also be present.

\bibitem{QCD_background} 
In each signature the QCD background distribution is derived from the
observed data distribution by using the background weight for each
observed event; the background level can thus be seen to follow the
data in the appropriate figures. The advantage of this procedure (as
opposed to just cutting on the track isolation variable) for the low
statistics on the tails of the distribution is that one can get some
sense of the level of background on the tails of distributions from
rare fragmentations of jets that may be topology dependent.

\bibitem{decay_in_flight} 
A kaon that decays before the COT volume results in a muon whose
momentum is correctly measured; a kaon  that decays after the COT
is itself correctly measured.
These contributions are included in the total background estimate.

%xxx Andrei- is this correct as written now?

\bibitem{Sasha} 
We thank A. Paramonov for the method and the code.

%\bibitem{loginov_thesis} A. Loginov, Ph.D thesis, Institute for
%Theoretical and Experimental Physics, Moscow, Russia, September,
%2006. \\
%The K-factor $K_W$ applied to $\Wg$ MadGraph MC samples is $K_W
%= 1.36$ for $M_W \le 76~\GeV$ and $K_W = 1.62 + 10^{-5}\times
%P_T^\gamma - 0.386\times e^{-0.1\times P_T^\gamma}$ for $M_W \gt
%76~\GeV$.  The K-factor $K_Z$ applied to $\Zg$ MadGraph MC samples is
%$K_Z = 1.36$ for $M_Z \le 86~\GeV$ and $K_Z = 1.46 - 0.000728\times
%P_T^\gamma- 0.125\times e^{-0.0615\times P_T^\gamma}$ for $M_Z \gt
%86~\GeV$.


\end{thebibliography}
\end{document}